\documentstyle[12pt,twoside, epsf, here, psfig]{article}
\newcounter{tony}
\newcommand{\bb}{\mbox{\boldmath{$b$}}}
\newcommand{\bc}{\mbox{\boldmath{$c$}}}

\newcommand{\be}{\mbox{\boldmath{$e$}}}
\newcommand{\fb}{\mbox{\boldmath{$f$}}}

\newcommand{\bn}{\mbox{\boldmath{$n$}}}

\newcommand{\bq}{\mbox{\boldmath{$q$}}}
\newcommand{\br}{\mbox{\boldmath{$r$}}}

\newcommand{\bt}{\mbox{\boldmath{$t$}}}

\newcommand{\bv}{\mbox{\boldmath{$v$}}}
\newcommand{\bw}{\mbox{\boldmath{$w$}}}
\newcommand{\bx}{\mbox{\boldmath{$x$}}}

\newcommand{\bA}{\mbox{\boldmath{$A$}}}
\newcommand{\bB}{\mbox{\boldmath{$B$}}}

\newcommand{\bD}{\mbox{\boldmath{$D$}}}

\newcommand{\bF}{\mbox{\boldmath{$F$}}}

\newcommand{\bH}{\mbox{\boldmath{$H$}}}
\newcommand{\bI}{\mbox{\boldmath{$I$}}}
\newcommand{\bJ}{\mbox{\boldmath{$J$}}}

\newcommand{\bN}{\mbox{\boldmath{$N$}}}

\newcommand{\bP}{\mbox{\boldmath{$P$}}}

\newcommand{\bT}{\mbox{\boldmath{$T$}}}

\newcommand{\bV}{\mbox{\boldmath{$V$}}}

\newcommand{\bepsilon}{\mbox{\boldmath{$\epsilon$}}}

\newcommand{\btheta}{\mbox{\boldmath{$\theta$}}}

\newcommand{\bxi}{\mbox{\boldmath{$\xi$}}}

\newcommand{\bsigma}{\mbox{\boldmath{$\sigma$}}}

\newcommand{\bvarphi}{\mbox{\boldmath{$\varphi$}}}

\newcommand{\bomega}{\mbox{\boldmath{$\omega$}}}
\newcommand{\assemb}{\mathop{\mbox{\LARGE\bf\sf A}}}

\newcommand{\beq}{\begin{equation}}
\newcommand{\eeq}[1]{\label{eq:#1}\end{equation}}

\newcommand{\eqref}[1]{(\ref{eq:#1})}

      \newcommand{\beqn}{\begin{equation}}
      \newcommand{\eeqn}{\end{equation}}
      \newcommand{\beqna}{\begin{eqnarray}}
      \newcommand{\eeqna}{\end{eqnarray}}

\newtheorem{theorem}{\sc Theorem}
\newtheorem{corollary}{\sc Corollary}

\title{Finite Element Simulation of the Motion of a Rigid Body in a
Fluid with Free Surface }

\renewcommand{\thefootnote}{\fnsymbol{footnote}}

\author{S. J. Childs \\ {\small\em Department of Pure and Applied
Mathematics, Rhodes University, Grahamstown,} \\ {\small\em 6140, South
Africa} \\ \\ B. D. Reddy \\ {\small\em Department of Mathematics and
Applied Mathematics, University of Cape Town,} \\ {\small\em
Rondebosch, 7700, South Africa}}

\date{}       

\begin{document}

\maketitle
\renewcommand{\thefootnote}{\arabic{footnote}}

\begin{abstract}
\noindent {\em In this work a finite element simulation of the motion
of a rigid body in a fluid, with free surface, is described. A
completely general referential description (of which both Lagrangian
and Eulerian descriptions are special cases) of an incompressible,
Newtonian fluid is used. Such a description enables a distorting
finite element mesh to be used for the deforming fluid domain. \\

\noindent A new scheme for the linearised approximation of the
convective term is proposed and the improved, second order accuracy
of this scheme is proved. A second theorem, which provides a
guideline for the artificial adjustment of the Reynolds number when
applying a continuation technique, is also proved. The most effective
means of eliminating pressure as a variable and enforcing
incompressibility are reviewed. A somewhat novel method to generate
finite element meshes automatically about included rigid bodies, and
which involves finite element mappings, is described. \\

\noindent The approach taken when approximating the free surface, is
that it may be treated as a material entity, that is, the material
derivative of the free surface is assumed zero. Euler's equations and
conservation of linear momentum are used to determine the motion of
the rigid body. A predictor--corrector method is used to solve the
combined sub--problems. The resulting model is tested in the context
of a driven cavity flow, a driven cavity flow with various, included
rigid bodies, a die--swell problem, and a Stokes second order wave.}
\end{abstract}


\section{Introduction}

\begin{figure}[H]
\begin{center} \leavevmode
\mbox{\epsfbox{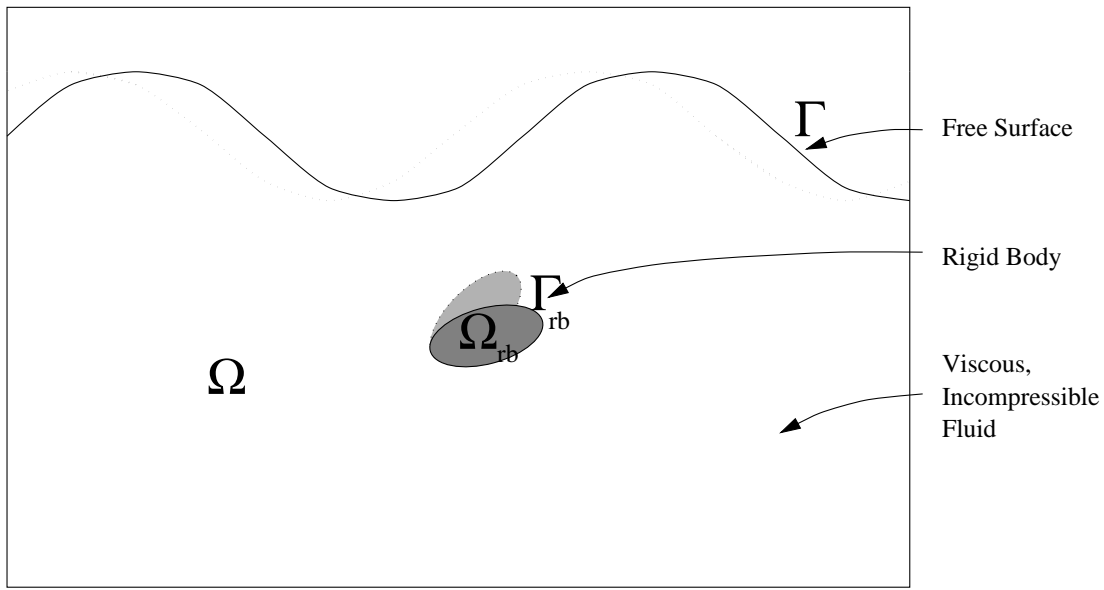}}
\end{center}
\caption{The Motion of a Rigid Body in a Fluid with Free Surface}
\label{143} 
\end{figure}

An enormous diversity of problems subscribe to the basic rigid
body--fluid--free surface theme (see Figure \ref{143} for a schematic
form of the problem of interest). The implications of free surfaces
and fluid--rigid body interactions are a deforming fluid domain and
the implementation of the majority of numerical time integration
schemes then becomes problematic where a purely Eulerian description
of the fluid motion has been used. The reason is that most numerical
time integration schemes require successive function evaluation at
fixed spatial locations.  Meshes rapidly snarl when purely Lagrangian
descriptions are used.

One alternative is to use the completely general referential
description proposed in {\sc Childs} \cite{me:1}.  Eulerian and
Lagrangian references are just two, specific examples of an infinite
number of configurations over which to define the fields used to
describe the dynamics of deforming continua. They are both special
cases of a more general referential description, a description in which
a reference configuration is chosen at will (closely related to the
so--called arbitrary Lagrangian Eulerian method or A.L.E.  method). The
completely general referential description is implemented in this
work.

Solving for the orientation and position of a rigid body entails
solving equations of motion. These are derived from the principles of
conservation of angular and linear momentum. Expressing the principle
of conservation of angular momentum in more usable terms is not as
straight forward as the linear momentum case, and a more appropriate
reference must be resorted to. This reference is fixed within the rigid
body in such a way that the origin and centre of mass coincide, as do
the axes and principal moments of inertia. A complete transformation of
all quantities, to this more appropriate reference, is used. This is in
contrast to the approach used in deriving the fluid equations, where
quantities are not perceived in terms of the new reference, but instead
merely defined to be functions of it. The change of reference is
accomplished using a transition matrix and the Coriolis theorem.

The approach taken when approximating the free surface, is that it may
be treated as a material entity i.e. the material derivative of the
free surface is assumed zero. Surface tension is not considered
significant in the free surfaces investigated. 

A predictor--corrector method is used to solve the combined rigid
body, fluid and free surface sub--problems. A backward difference
scheme is used to approximate the time derivative in the fluid
sub--problem (in this regard see {\sc Childs} \cite{me:1}). The
finite element method is used for the spatial (referential ``space'')
discretisation. Nonlinearity is circumvented by way of a new second
order accurate linearisation, and the use of Picard iteration if
necessary. Picard iteration amounts to ``linearising'' the
non--linear term with the solution obtained during the previous
iterate. A penalty method is used to eliminate pressure as a variable
and a $Q_2$--$P_1$ element pair is used as a basis. A somewhat novel
method to generate finite element meshes automatically about included
rigid bodies, and which involves finite element mappings, is
described. 

The more conventional methods for solving initial value problems are
used to solve for the position and orientation of the rigid body. The
definitions of linear and angular velocity are coupled to the
equations of motion to produce a system of twelve equations (which
are of course also coupled to the fluid equations) and a
Runge--Kutta--Fehlberg method is then used. A backward difference
scheme and the finite element method are used for the respective
temporal and spatial discretisations in the free surface
sub--problem.

The overall model was tested in the context of a driven cavity flow, a
driven cavity flow with various, included rigid bodies, a die--swell
problem and a Stokes, second--order wave. Results for these examples
are presented in Section \ref{4}. The programme was written by the
author in Fortran and originally run on a model 400 Dec Alpha 3100.

\section{The Equations Governing the Motion of the Rigid Body}

What follows is a brief exposition of the equations of motion. This is,
of course, review of classical work contained in a great variety of
references. Its inclusion is deemed appropriate not for the sake of
completeness alone, but because Euler's equations (the equations
governing the angular motion) and the equations governing the
translation are written in terms of two distinctly different
references. 

The equations of motion, six in all, are essentially an explicit
re--write of the principles of linear and angular momentum.  The
equations which govern the translation,
\[
\frac{d {\bv}_{rb}}{dt} = \frac{1}{m} \Sigma {\fb},
\]
need no introduction, they being a straight forward statement of the
principle of conservation of linear momentum. In the above equation
${\bv}_{rb}$ denotes the velocity of the rigid body, $m$ denotes the
mass, $\Sigma {\fb}$ is the sum of the external forces acting on the
rigid body and $t$ denotes time. Expressing the angular momentum
equations in a more practicable form is not as straight forward and
necessitates the change to the reference already mentioned. The
conservation principle of angular momentum may be stated as \renewcommand{\thefootnote}{\fnsymbol{footnote}}
\begin{eqnarray} \label{24}
\frac{d {\hat {\bq} }}{dt} &=& \Sigma ({\hat {\br} } \wedge {\hat
{\fb} }) \footnotemark[2]
\end{eqnarray}
where ${\hat {\bq} }$ is the angular momentum and $\Sigma ({\hat {\br}}
\wedge {\hat {\fb}})$ is the sum of the torques which arise as a result
of the external forces acting about the centre of mass (${\hat {\br}}$
is the position vector of a material point within the rigid body and
${\hat {\fb}}$ is the force). The equations for the conservation of
angular momentum, though easily understood, are not of much use when
stated in the above form. \footnotetext[2]{The \ ${\hat {} \ }$s are
resorted to temporarily in anticipation of the change to the new
reference.} \renewcommand{\thefootnote}{\arabic{footnote}}

\subsubsection*{A More Appropriate Reference}

\begin{figure}[H]
\begin{center} \leavevmode
\mbox{\epsfbox{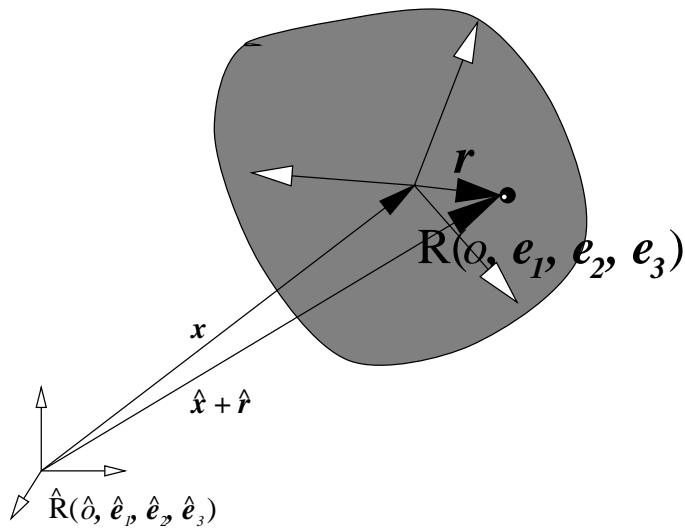}}
\end{center}
\caption{Schematic diagram depicting the position vectors ${\bx}$ and
${\br}$ in terms of a more appropriate reference,
$R(o,{\be}_1,{\be}_2,{\be}_3)$, which is embedded in the rigid body in
such a way that the origin, $o$, and centre of mass coincide as do the
basis vectors and the principal moments of inertia.} \label{101}
\end{figure}

The fact that 
\[
\int_{\Omega_{rb}} \rho {\br}  \ d{\Omega_{rb}} = {\bf 0}
\]
for a reference whose origin coincides with the centre of mass
(${\Omega_{rb}}$ is the rigid body domain depicted in Fig. \ref{101}) immediately suggests the
existence of a more appropriate reference than the one in terms of
which the above conservation principle is written. For such a
reference fixed within the rigid body,
\[
\frac{d {\br} }{dt} = {\bf 0}
\]
in addition. This suggests expressing 
\begin{eqnarray*}
{\hat {\bq}} &=& \int_{\Omega_{rb}} \rho {\hat {\br}} \wedge {\hat
{\bv} } \ d{\hat {\Omega}}_{rb} \\
&=& \int_{\Omega_{rb}} \rho {\hat {\br}} \wedge \frac{d{\hat {\bf
r}}}{dt} \ d{\hat {\Omega}}_{rb} + \int_{\Omega_{rb}} \rho {\hat {\br}
} \ d{\hat {\Omega}}_{rb} \wedge \frac{d{\hat {\bx} }}{dt}
\end{eqnarray*}
in terms of a more appropriate reference. Converting to this new, more
appropriate reference by means of the Coriolis theorem, the relation
\begin{eqnarray*}
{\bq}  &=& \int_{\Omega_{rb}} \rho {\br}  \wedge ( {\bomega} \wedge
{\br} ) d{\Omega_{rb}} \hspace{10mm} \left( \displaystyle \frac{d {\br}
}{dt} = {\bf 0} \hspace{5mm} \mbox{\it for the new reference} \right)
\end{eqnarray*}
is obtained, where ${\bomega}$ is the angular velocity. It is then a
fairly elementary exercise to show that
\begin{eqnarray} \label{20}
{\bq} = {\bJ}{\bomega},
\end{eqnarray}
where ${\bJ}$ is the inertia matrix defined by 
\[
J_{ij} = \int_{\Omega_{rb}} \rho ( r_k r_k \delta_{ij} - r_i r_j )
d{\Omega_{rb}}.
\]
This result is demonstrated in {\sc Woodhouse} \cite{w:1}.
Exploring this idea of a more appropriate reference further and
restricting it to one, final choice allows still greater
simplification; by specifying the basis vectors to coincide with the
eigenvectors of the inertia matrix ${\bJ}(R)$ (i.e. the basis vectors
are principal axes), equation (\ref{20}) reduces to
\begin{eqnarray} \label{23}
{\bq} &=& \left[ \begin{array}{c} J_{11} \omega_1 \\ J_{22} \omega_2
\\ J_{33} \omega_3 \end{array} \right].
\end{eqnarray}

\subsubsection*{Conservation of Angular Momentum in Terms of this More
Appropriate Reference}

Rewriting equation (\ref{24}) in terms of this new reference leads
firstly to
\begin{eqnarray*}
\frac{d {\bq} }{dt} + {\bomega} \wedge {\bq}  &=&
\Sigma ({\br}  \wedge {\fb}) \hspace{10mm} \mbox{\it (by
the Coriolis theorem),} 
\end{eqnarray*}
which in terms of the relation (\ref{23}), is
\begin{eqnarray*}
J_{11} \frac{d \omega_{1}}{dt} + (J_{33} - J_{22}) \omega_2 \omega_3
&=& \Sigma ({\br} \wedge {\fb}) \cdot {\be}_1 \\
J_{22} \frac{d \omega_{2}}{dt} + (J_{11} - J_{33}) \omega_3 \omega_1
&=& \Sigma ({\br} \wedge {\fb}) \cdot {\be}_2 \\
J_{33} \frac{d \omega_{3}}{dt} + (J_{22} - J_{11}) \omega_1 \omega_2
&=& \Sigma ({\br} \wedge {\fb}) \cdot {\be}_3
\end{eqnarray*}
where the $J_{ii \mbox{\scriptsize \ (no contraction)}}$ are the
principal moments of inertia defined by
\begin{eqnarray*}
\int_{\Omega_{rb}} \rho r^2_i d \Omega_{rb}.
\end{eqnarray*}
This description is in terms of a reference embedded within the rigid
body in such a way that the origin and centre of mass coincide, as do
the axes and principal moments of inertia. Failure to recognise this
can lead to an incorrect formulation of the forces acting on the rigid
body.

\section{A Free Surface Description Ignoring Surface Tension}
\label{37}

It is assumed that the vertical position of the free surface can be
written as a function of horizontal coordinates and time, that is, $x_3
= h(x_1,x_2,t)$. The free surface can be described equivalently by the
condition
\[ 
\eta({\bx} ,t) \equiv h(x_1,x_2,t) - x_3 = 0.
 \]
Requiring that the material derivative of this free surface to be zero
amounts to writing
\begin{eqnarray} \label{113} 
\frac{\partial h}{\partial t} + {\nabla h} \cdot [ v_1, v_2 ] = v_3. 
\end{eqnarray}
A physical interpretation of this mathematics is that the free surface
is an infinitesimally thin, perfectly elastic skin by which the fluid
is contained. The free surface is in actuality not a distinct material
entity. The approach is consequently slightly limited. For example, no
convection of the free surface is allowed. Restricting the movement of
surface nodes to vertical translations alone circumvents having to
rewrite the above equations in terms of a completely general
referential description. 

The variational equations are obtained in the usual manner. Multiplying
by an arbitrary function $w$ and integrating, the equations
\begin{eqnarray} \label{157}
\int_{\Gamma} w \frac{\partial h}{\partial t}{d \Gamma} \ +
\ \int_{\Gamma} w {\nabla h} \cdot [ v_1, v_2 ] {d \Gamma} &=&
\int_{\Gamma} w v_3 {d \Gamma} \nonumber
\end{eqnarray}
are obtained. 

Equation (\ref{113}) is hyperbolic and one would therefore anticipate
discontinuities or shocks in the solution. A standard Galerkin type
formulation is not sufficient (see {\sc Hughes} \cite{h:3}, {\sc
Hughes, Mallet} and {\sc Mizukami} \cite{h:4} and {\sc Johnson}
\cite{j:1}) under such circumstances.

The Streamline--Upwind--Petrov--Galerkin (S.U.P.G.) Method with
discontinuity capturing is one popular method resorted to when solving
hyperbolic equations. This amounts to using
\[ 
w = w^{\ast} \ + \ {\tau_1}({\nabla w^{\ast}} \cdot {\bv}) \ +
\ {\tau_{2}}({\nabla w^{\ast}} \cdot {\bv} _{tangent}), \hspace{10mm}
{\bv}_{tangent} = \frac{ {\bv} \cdot {\nabla h} }{ \mid\mid {\nabla h}
\mid\mid^2 } {\nabla h},
\]
where $w^{\ast}$ is the arbitrary function in the formulation, $\tau_1$
and $\tau_2$ are constants described in {\sc Hughes}, {\sc Mallet} and
{\sc Mizukami} \cite{h:4}.

Hermite elements are one more way in which to obtain smoother
surfaces.  Hermite elements are used for $C^1$ approximations which
require non--zero first order derivatives at the nodes.

{\sc Remark:} Although surface tension is not considered significant in
the free surfaces to be described in this work, the free surface
equations to be used under such circumstances would be
\begin{eqnarray*}
({\bsigma}_1 - {\bsigma}_2) \cdot \frac{\nabla \eta}{\mid\mid {\nabla
\eta} \mid\mid} = 2 G \zeta  \frac{\nabla \eta}{\mid\mid {\nabla \eta}
\mid\mid} + {\nabla \zeta}
\end{eqnarray*}
where the left hand side is the total traction (the difference between
that acting from within and that acting from without) acting on the
interface, the last term is a tangential force contribution in
instances where surface tension is non-constant, $G$ contains the mean
curvature
\begin{eqnarray*}
\frac{\partial}{\partial x_1} \left\{ \frac{({\nabla \eta})_1 +
({\nabla \eta})_2}{\mid\mid {\nabla \eta} \mid\mid} \right\} +
\frac{\partial}{\partial x_2} \left\{ \frac{({\nabla \eta})_1 +
({\nabla \eta})_2}{\mid\mid {\nabla \eta} \mid\mid} \right\},
\end{eqnarray*}
and $\zeta$ is the surface tension. See {\sc Landau} and {\sc Lifshitz}
\cite{l:1}, {\sc Bird}, {\sc Armstrong} and {\sc Hassager} \cite{b:4}
and {\sc Joseph} and {\sc Renardy} \cite{j:2} in this regard.

\section{A Completely General Referential Description of an
Incompressible, Newtonian Fluid}

The implications of free surfaces and fluid--rigid body interactions
are a deforming fluid domain and the implementation of the majority of
numerical time integration schemes then becomes problematic where a
purely Eulerian description of the fluid motion has been used. The
reason is that most numerical time integration schemes require
successive function evaluation at fixed spatial locations. Meshes
rapidly snarl when purely Lagrangian descriptions are used. One
alternative is to use a completely general referential description.

The transformation to the completely general reference only involves
coordinates where used as spatial variables and the resultant
description is therefore inertial in the same way as Lagrangian
descriptions are. The conservation principles of mass and linear
momentum remain the basis of the fluid model despite having resorted to
a completely general reference (as opposed to a conventional Eulerian
reference) for the description of the fluid. For the purposes of
reading the equations which folow some slightly specialised notation
will be introduced.

\subsubsection*{Notation} In the equations that follow the \ ${\tilde
{}} \ $s merely indicate that quantities are functions of a deforming
reference, ${\tilde {\bx}}$. The operators ${\tilde \nabla}$ and
$\widetilde {\mbox{div}}$ are the referential counterparts of $\nabla$
and $\mbox{div}$, that is to say that
\[ 
{\tilde \nabla} = \frac{\partial}{\partial {\tilde {\bx}}}
\hspace{10mm} \mbox{and} \hspace{10mm} {\widetilde {\mbox{div}}} =
\frac{\partial}{\partial {\tilde x}_1} + \frac{\partial}{\partial
{\tilde x}_2} + \frac{\partial}{\partial {\tilde x}_3}.
\] 
The symbol ${\tilde J}$ is used to denote the determinant of the
deformation gradient
\[
{\bf \tilde{F}} = \frac{\partial {\bx}}{\partial {\tilde {\bx}}}
\]
and the notation ${\bA}:{\bB}$ is used to denote the matrix inner
product $A_{ij}B_{ij}$.

In terms of this newly established notation the conservation principles
of linear momentum and mass may be written in primitive form as
\begin{eqnarray} \label{29}
{\rho} \left[ \frac{\partial {\tilde {\bv} }}{\partial t} + {{{\tilde
\nabla} \tilde {\bv} } {\tilde {\bF} }^{-1} } ({\tilde {\bv} } -
{\tilde {\bv} }^{ref}) \right] {\tilde J} &=& {\rho} {\tilde {\bb} }
{\tilde J} + \mathop{\widetilde {\rm div}}{\tilde {\bP}} \nonumber
\end{eqnarray}
and
\begin{eqnarray} \label{30}
{{\tilde \nabla} {\tilde {\bv}} : {\tilde {\bF} }^{-t} } \ = \ 0,
\nonumber
\end{eqnarray}
where $\tilde{\bP}$ is the Piola--Kirchoff stress tensor of the first
kind, $\tilde{\bP}  \ = \ \tilde{\bsigma} {\tilde {\bF}}^{-t} {\tilde
J}$. In terms of the constitutive relation for a Newtonian fluid,
$\bsigma = - p {\bI} + 2 \mu {\bD}$. Thus
\[
{\tilde {\bP}} = \left( - p{\bI} + \mu \left[ {\tilde \nabla}{\tilde
{\bv}}{\tilde {\bF}}^{-1} + \left({\tilde \nabla}{\tilde {\bv}}{\tilde
{\bF}}^{-1}\right)^t \right] \right){\tilde {\bF}}^{-t} {\tilde J}
\hspace{8mm} \mbox{since} \hspace{8mm} {\tilde {\bD}} = \frac{1}{2}
\left[ {\widetilde {\nabla {\bv}}} + \left( {\widetilde{\nabla {\bv}}}
\right)^t \right].
\]
A comprehensive derivation of these equations is given in {\sc Childs}
\cite{me:1}. The derivation of a variational formulation is along
similar lines to that for the conventional Navier-Stokes equations (the
purely Eulerian description).

For a fluid of constant density, the variational formulation
\begin{eqnarray} \label{33}
{\rho} \int_{ \tilde \Omega } {\tilde {\bw} } \cdot \frac{\partial
{\tilde {\bv} }}{\partial t} {\tilde J} {d{\tilde \Omega}} \ + \ {\rho}
\int_{ \tilde \Omega } {\tilde {\bw}} \cdot {{\tilde \nabla} {\tilde
{\bv} }} \left[{\tilde {\bF} }^{-1} ({\tilde {\bv}} - {\tilde {\bv}
}^{ref}) \right] {{\tilde J}}{d {\tilde \Omega}} \ = \nonumber
\hspace{40mm} & & \\
{\rho} \int_{ \tilde \Omega } {\tilde {\bw} } \cdot {\tilde
{\bb}} {\tilde J} {d {\tilde \Omega}} \ + \ \int_{
\tilde \Omega } {\tilde p} {{{\tilde \nabla} \tilde {\bw} } : {\tilde
{\bF} }^{-t}} {\tilde J} {d {\tilde \Omega}} \ - \ 2
{\mu} \int_{ \tilde \Omega } {\tilde {\bD} }(\tilde {\bw} ) :
{\tilde {\bf D}}(\tilde {\bv}) {\tilde J} {d {\tilde
\Omega}} \nonumber & & \\
+  {\rho} \int_{ \tilde \Gamma } {\tilde {\bw}} {\tilde {\bP}} {\tilde
{\bN}} d \Gamma \hspace{50mm} & &  \nonumber
\end{eqnarray}
\begin{eqnarray} \label{34}
\int_{\tilde \Omega} {\tilde q} { {{\tilde \nabla} \tilde {\bv} } :
{\tilde {\bF} }^{-t} } {d {\tilde \Omega}} &=& 0  \nonumber
\end{eqnarray}
is obtained, where $\tilde q$ and ${\tilde {\bw}}$ are respectively the
arbitrary pressure and velocity of the variational formulation. Under
the circumstances of a Dirichlet boundary condition the term involving
the boundary integral may naturally be omitted.

\section{Statement of the Total Problem in Dimensionless Form}

Suppose that length is to be measured in units of $X$ and velocity in
units of $V$ (which is really parameterising time by $T =
\frac{X}{V}$). Then 
\[
{\bx} = {\bar {\bx}} X, \hspace{10mm} {\bv} = {\bar {\bv}} V
\hspace{10mm} and \hspace{10mm} t = {\bar t} \frac{X}{V}.
\] 
The equations governing the translation of the rigid body are
accordingly divided through by $\frac{V^2}{X}$, the equations governing
the rotation by $\rho_s X^3 V^2$, and the fluid equations by $\frac{V^2
\rho}{X}$.

Using the symbol ${\bar J}_{ii \mbox{\scriptsize (no sum)}}$ to denote
the $i$th, dimensionless principal moment of inertia of the rigid body,
\[
{\bar J}_{ii\mbox{\scriptsize (no sum)}} = \frac{J_{ii\mbox{\scriptsize
(no sum)}}}{\rho_s X^5},
\]
${\bH}$ to denote the transition matrix for a transition to a reference
whose axes coincide with these principal moments of inertia, ${\bar {\bc}}$ to denote the centre of mass of the
rigid body, ${\bar \Gamma}_{rb}$
to denote the dimensionless surface of the rigid body, ${\bar t}$ to denote a
dimensionless time, $\rho_f$ and $\rho_s$ to denote the density of fluid and solid respectively, ${\bar h}$ to denote the
dimensionless elevation of a free surface (${\bar h}$ being a function
of a reference prohibited from deforming in a
lateral direction i.e. nodes belonging to the free surface move in the vertical only), ${\bar {\bv}}$ to denote the dimensionless velocity of the
underlying fluid, ${\bar {\bF}}$ to denote the deformation gradient, ${\bar
J}$ its determinant, ${\bar {\bv}}^{\scriptsize \mbox{\it mesh}}$ to denote
the dimensionless velocity of a mesh which is otherwise allowed to
deform freely and $Re$ to denote the Reynolds number,
\[
Re = \frac{XV \rho_f}{\mu},
\]
the combined, dimensionless, free surface--fluid--rigid body problem
can then be stated as follows: find ${\bar {\bx}}_{rb}$, $\bar
{\btheta}$, $\bar {\bv}_{rb}$, $\bar {\bomega}$, $\bar h$ and $\bar
{\bv}$ (the dimensionless respective position, orientation, velocity
and angular velocity of the rigid body, the elevation of the free
surface and the velocity field of the fluid), which satisfy
\
\begin{eqnarray*}
\begin{array}{cl}
\left. 
\begin{array}{c}
\begin{array}{l}
{\bar J}_{11} \displaystyle \frac{d {\bar \omega}_{1}}{d{\bar t}} \ +
\ ({\bar J}_{33} \ - \ {\bar J}_{22}) {\bar \omega}_2 {\bar \omega}_3 =
\\
\hspace{40mm} \displaystyle \frac{\rho_{\scriptsize
f}}{\rho_{\scriptsize s}} \left[ {\bH} \displaystyle \int_{\bar
\Gamma_{rb}} ({\bar {\bx}}-{\bar {\bc}} ) \wedge \left\{-{\bar p}{\bI}
+ \frac{2}{Re} {\bar {\bD}} \right\} {\bn} \ d{\bar \Gamma}_{rb}
\right] \cdot {\be}_1 \\
\\ 
{\bar J}_{22} \displaystyle \frac{d {\bar \omega}_{2}}{d{\bar t}} \ +
\ ({\bar J}_{11} \ - \ {\bar J}_{33}) {\bar \omega}_3 {\bar \omega}_1 =
\\
\hspace{40mm} \displaystyle \frac{\rho_{\scriptsize
f}}{\rho_{\scriptsize s}} \left[ {\bH} \displaystyle \int_{\bar
\Gamma_{rb}} ({\bar {\bx}}-{\bar {\bc}}) \wedge \left\{-{\bar p}{\bI}
+ \frac{2}{Re} {\bar {\bD}} \right\} {\bn} \ d{\bar \Gamma}_{rb}
\right] \cdot {\be}_2 \\
\\
{\bar J}_{33} \displaystyle \frac{d {\bar \omega}_{3}}{d{\bar t}} \ +
\ ({\bar J}_{22} \ - \ {\bar J}_{11}) {\bar \omega}_1 {\bar \omega}_2 =
\\
\hspace{40mm} \displaystyle \frac{\rho_{\scriptsize
f}}{\rho_{\scriptsize s}} \left[ {\bH} \displaystyle \int_{\bar
\Gamma_{rb}} ({\bar {\bx}}-{\bar {\bc}}) \wedge \left\{-{\bar p}{\bI}
+ \frac{2}{Re} {\bar {\bD}} \right\} {\bn} \ d{\bar \Gamma}_{rb}
\right]\cdot {\be}_3 \end{array} \\
\\ 
\begin{array}{ccc} \displaystyle \frac{d{\bar {\bv}}_{rb}}{d{\bar t}}
&=& \displaystyle \frac{\rho_{\scriptsize f}}{{\bar m}
\rho_{\scriptsize s}} \int_{\bar \Gamma_{rb}} \left\{ - {\bar p}{\bI} +
\frac{2}{Re} {\bar {\bD}} \right\} {\bn} \ d{\bar \Gamma}_{rb} +
\frac{X}{V^2}{\bb} \end{array} \\
\\
\begin{array}{ccc} \displaystyle \frac{d {\bar {\btheta}}}{dt} &=&
{\bar {\bomega}} \end{array} \\
\\
\begin{array}{ccc} \displaystyle \frac{d {\bar {\bx}}_{rb}}{dt} &=&
{\bar {\bv}}_{rb} \end{array}
\end{array} 
\right\}
&
\begin{array}{l} 
\mbox{Included} \\ \mbox{Rigid} \\ \mbox{Body}
\end{array} \\
& \\
& \\
& \\
\left. \displaystyle \frac{\partial {\bar h}}{\partial {\bar t}} +
{{\bar \nabla} {\bar h}} \cdot [{\bar v}_1, {\bar v}_2 ] = {\bar v}_3
\hspace{10mm} \right\}
& \begin{array}{l} \mbox{Free} \\ \mbox{Surface} \end{array} \\
& \\
& \\
& \\
\left. \begin{array}{lc} & \begin{array}{rcl} \left[ \displaystyle
\frac{\partial {\bar {\bv} }}{\partial {\bar t}} + {{{\bar \nabla} \bar
{\bv} } {\bar {\bF} }^{-1} } ({\bar {\bv} } - {\bar {\bv} }^{mesh})
\right] {\bar J} &=& \displaystyle \frac{X}{V^2}{\bb} {\bar J} +
\mathop{\bar {\rm div}}{\bar {\bP}} \\
& & \\
{ {{\bar \nabla} \bar {\bv} } : {\bar {\bF} }^{-t} } & = & 0
\end{array} \\
& \\
& \mbox{where} \hspace{100mm} \\ 
& \\
&  {\bar {\bP}}  \ = \ {\bar {\bsigma}} {\bar {\bF}}^{-t}  {\bar J} \ =
\ \left( - {\bar p}{\bI} + \displaystyle \frac{2}{\mbox{\it Re}} \left[
{\bar \nabla}{\bar {\bv}}{\bar {\bF}}^{-1} + \left({\bar \nabla}{\bar
{\bv}}{\bar {\bF}}^{-1} \right)^t \right] \right) {\bar {\bF}}^{-t}
{\bar J}
\end{array}
\right\}
& 
\begin{array}{l} \mbox{Fluid} \end{array}
\end{array} \\
\end{eqnarray*}
\ 
subject to the ``no slip'' requirements 
\begin{eqnarray*}
{\bar {\bv}} \mid_{{\bar \Gamma}_{rb}} &=& {\bar {\bv}}_{rb} + {\bar
{\bomega}} \wedge ({\bar {\bx}} - {\bar {\bc}})
\end{eqnarray*}
at fluid--rigid body interfaces and
\begin{eqnarray*}
{\bar {\bv}} \mid_{\bar \Gamma} &=& {\bf 0} 
\end{eqnarray*}
at fixed, solid impermeable boundaries. Additional boundary conditions
depend on the problem in question.

\section{Approximate Fluid Equation and its Implementation}

Two topics which are relevant to the approximate fluid equation are
dealing with the nonlinearity of the convective term and enforcing
incompressibility in such a way as to eliminate pressure as a variable.
Only against such a background can the approximate fluid equation be
written.

\subsection{The Incompressibility Condition}

Two methods to enforce incompressibility and eliminate pressure as a
variable are considered relevant for review. The use of either a
penalty method or an iterative, augmented Lagrangian method both have
distinct, yet different advantages. 

The L.B.B. or B.B. condition (so named after Ladyzhenskaya, Babu\v{s}ka
and Brezzi) must be taken into account in the implementation of either
method. So--called ``locking'' or ``chequerboard'' modes can arise in
the event of the L.B.B. condition not being satisfied. Work by authors
such as {\sc Oden, Kikuchi} and {\sc Song} \cite{oden:1} can be
referred to for a more in-depth treatment of the L.B.B. condition.

A penalty method was used to eliminate pressure as a variable and
enforce incompressibility in later examples, despite the acclaimed
superiority of the iterative augmented Lagrangian approach (discussed
shortly). The results obtained using the more efficient and more easily
implemented penalty method compared favourably with those obtained by
{\sc Simo} and {\sc Armero} \cite{s:1}, who used the iterative
augmented Lagrangian approach. In other examples the penalty method was
used for reasons of standardisation (so that results could be compared
against those cited in the literature). 

\subsubsection*{The Simple Penalty Method}

Penalty methods are closely related to the artificial compressibility
method and provide a remarkably simple strategy for solving the
equations. A fundamental assumption made in using the penalty method is
that the incompressibility condition,
\[
\mathop{ {\rm div}} { {\bv}} = 0,
\]
can be replaced by one of small incompressibility,
\[
\mathop{ {\rm div}} { {\bv}}  = - \epsilon {p},
\]
where $\epsilon$ is small. (It is tempting to substitute this
expression for $p$ in the momentum equations immediately, thereby
eliminating the pressure from the equations altogether at an analytic
stage and prior to construction. This would circumvent the complicated
construction which otherwise results. If such an approach were to be
taken, however, the L.B.B. condition would not be satisfied and an
appropriate underintegration rule would then be necessary to avoid an
approximation which will lead to ``locking'' or ``chequer board''
modes.)

The corresponding variational, small incompressibility is
\begin{eqnarray} \label{72}
\int_{ \Omega} { q} \mathop{ {\rm div}} { {\bv}} d { \Omega} = -
\epsilon \int_{ \Omega} { q} { p} d{ \Omega} \nonumber
\end{eqnarray}
where $q$ is the arbitrary function of the variational formulation. A
well known shortcoming of penalty methods is the severe
ill--conditioning which occurs as the penalty parameter is decreased,
making the simulation of exact incompressibility practically
impossible. As the theoretical limit $\epsilon = 0$, an infinite
condition number for the discrete algebraic system is obtained. Despite
this shortcoming a straight forward penalty method provides a
remarkably easy--to--implement means of enforcing incompressibility and
eliminating pressure as a variable, furthermore, there was no
discernable difference between results obtained using the penalty
method and those obtained using the iterative augmented Lagrangian
approach for the driven cavity flow problem (compare the results in
Subsection \ref{2} with those obtained by {\sc Simo} and {\sc Armero}
\cite{s:1}).

\subsubsection*{An Iterative Augmented Lagrangian Approach for Enforcing
Exact Incompressibility} \label{63}

There is considerable merit in the iterative augmented Lagrangian
approach from the point of view of realising theory. Incompressibility
can be exactly enforced (to within a specified tolerance) for finite
values of the penalty parameter using an iterative augmented Lagrangian
approach. This is in contrast to conventional penalty formulations
where there is a trade--off between enforcing exact incompressibility
and obtaining a severely ill--conditioned algebraic system.

A fundamental assumption made in using the iterative augmented
Lagrangian method is that the incompressibility condition can be
replaced by one of iterative, limiting--case incompressibility,
\[
\mathop{ {\rm div}} { {\bv}}  = - \epsilon \left(
{ p}^{(k+1)} -  { p}^{(k)} \right)
\]
for $\epsilon$ small. 

The method of augmented Lagrangians provides an effective means to
circumvent the shortcomings of the simple penalty method and enables
the divergence free constraint to be exactly enforced for finite values
of the penalty parameter.

\subsection{A New, Linear Scheme for the Approximation of the
Convective Term} \label{1}

The following theorem is the first of two proposed numerical
improvements (originally presented in {\sc Childs} and {\sc Reddy}
\cite{me:3}). Linearising with a guess obtained by extrapolating
through solutions from the previous two time steps leads to second
order accuracy. This is an improvement on the conventional method by
an order of magnitude.

\begin{theorem} \label{148}
The linearised terms, $ ( 2 {\bv} \mid_t - {\bv} \mid_{t-\Delta t} )
\cdot \nabla {\bv} \mid_{t+\Delta t} $ and $ {\bv} \mid_{t+\Delta t}
\cdot \nabla (2 {\bv} \mid_t - {\bv} \mid_{t-\Delta t})$, are second
order accurate (have error $ O(\Delta t^2)$) approximations of the
nonlinear term $({\bv} \cdot \nabla {\bv}) \mid_{t+\Delta t}$.
\end{theorem}
{\sc Proof:} 
\begin{eqnarray*}
{\bv} \mid_{t+\Delta t} &=& {\bv} \mid_t + \Delta t \left.
\frac{\partial {\bv}}{\partial t} \right|_{t} + O(\Delta t^2)
\hspace{10mm} \mbox{\it (by Taylor series)} \\
&=& {\bv} \mid_{t} + \Delta t \left( \frac{ {\bv} \mid_t - {\bv}
\mid_{t-\Delta t} }{ \Delta t } + O(\Delta t) \right) + O(\Delta t^2)
\hspace{10mm}  \mbox{\it (using a backward} \\
& & \hspace{90mm} \mbox{\it difference)} \\
&=& 2{\bv}\mid_t - {\bv}\mid_{t-\Delta t} + O(\Delta t^2) \\
&& \\
\left( {\bv} \cdot \nabla {\bv} \right) \mid_{t+\Delta t} &=& \left[
2{\bv}\mid_t - {\bv}\mid_{t-\Delta t} + O(\Delta t^2) \right]
\frac{}{} \cdot {\nabla} {\bv} \mid_{t+\Delta t} \\
&=& \left[ 2{\bv}\mid_t - {\bv}\mid_{t-\Delta t} \right] \cdot \nabla
{\bv} \mid_{t+\Delta t} + O(\Delta t^2) \frac{}{}
\end{eqnarray*}
\renewcommand{\thefootnote}{\fnsymbol{footnote}}
The above linearisation schemes are an improvement on the conventional
${\bv}\mid_t \cdot \nabla {\bv} \mid_{t+\Delta t} \footnotemark[2]$
\footnotetext[2]{Favoured in terms of both rate and radius of
convergence by {\sc Cuvelier}, {\sc Segal} and {\sc van Steenhoven}
\cite{c:3}.} or ${\bv} \mid_{t+\Delta t} \cdot \nabla {\bv} \mid_t$
linearisation schemes by an order of magnitude.
\renewcommand{\thefootnote}{\arabic{footnote}}  
 
Further refinement of this second order accurate solution can usually
be accomplished by the process of Picard iteration, if necessary --
providing the initial guess lies within a radius of convergence. Picard
iteration amounts to ``linearising'' the nonlinear term with the
solution obtained during the previous iterate.

\subsection{Approximate Fluid Equation}

An updated approach and the choice of a referential configuration which
coincides with the spatial configuration at the instant within each
time step about which the equations are to be evaluated, facilitates
the elimination of the deformation gradient from the resulting fluid
scheme completely (under such conditions the deformation gradient
becomes identity -- see {\sc Childs} \cite{me:1}).

A backward difference is used to approximate the time derivative in
accordance with the findings of {\sc Childs} \cite{me:1}. The finite
element method is used for the spatial (referential ``space'' -- the
reference deforms) discretisation and a $Q_2$--$P_1$ element pair is
the chosen basis. A penalty method is used to eliminate pressure as a
variable.  The approximate equations, to be solved for the nodal
velocities, $V^e_j$ (pertaining to element $e$), are consequently
\begin{eqnarray*}
\assemb_{e=1}^E \ \left\{ \ \frac{1}{\Delta t} \int_{\hat{\Omega}}
B_{ki} B_{kj} J^e {d{\hat{\Omega}}} \right. &+& \int_{\hat{\Omega}}
B_{ki} G_{kjl} B_{lm} J^e {d{\hat{\Omega}}} (V_m^{e \ linearisation} -
V_m^{e \ mesh}) \\
&& \\
&+& \ \frac{1}{{\epsilon}} \left[ \int_{\hat{\Omega}}
{{\varphi}_m} {{\varphi}_l} J^e {d{\hat{\Omega}}} \right]^{-1}
\int_{\hat{\Omega}} {{\varphi}_m} G_{kjk} J^e {d{\hat{\Omega}}}
\int_{\hat{\Omega}} {{\varphi}_l} G_{nin} J^e {d{\hat{\Omega}}} \\
&& \\
&+& \ \left. \frac{2}{Re} \int_{\hat{\Omega}} \left(
G_{kil} + G_{lik} \right) \left( G_{kjl} + G_{ljk} \right) J^e
{d{\hat{\Omega}}} \ \right\} \assemb_{e=1}^E V^e_j 
\end{eqnarray*}
\[ 
= \ \assemb_{e=1}^E \ \left\{ \int_{\hat{\Omega}} g \ B_{ki} B_{k3} J^e
{d{\hat{\Omega}}} \ + \ \frac{1}{{\Delta}t} \int_{\hat{\Omega}} B_{ki}
B_{kj} J^e {d{\hat{\Omega}}} \left( V^e_j{\mid}_{t - {\Delta}t} \right)
\right\},
\]
where $\assemb$ is the element assembly operator, $E$ is the total
number of elements, $e$, into which the domain has been subdivided,
$\hat{\Omega}$ is the master element domain, $\Delta t$ is the length
of the time step, the $\phi_i({\bxi})$ are the basis functions, $\{
{\bxi} \}$ is the master element coordinate system,
\renewcommand{\thefootnote}{\fnsymbol{footnote}}
\[ 
{\bB} ({\bxi}) = { \left[ \begin{array}{lllllll}
{\phi_1} & 0 & 0 & \ldots & \phi_n & 0 & 0 \\ 0 & \phi_1 & 0 & \ldots &
0 & \phi_n & 0 \\ 0 & 0 & \phi_1 & \ldots & 0 & 0 & \phi_n \end{array}
\right] } \ , \hspace{10mm}
{\bvarphi}(\bxi) = \left[ 1,
x_1(\bxi), x_2(\bxi),
x_3(\bxi) \right] \footnotemark[2] \ , 
\] 
\[ 
G_{ikj}({\bxi}) = \frac{\partial B_{ik}}{\partial{\xi}_l}
\frac{{\xi}_l}{\partial { x}_j} = \frac{\partial B_{ik}}{\partial {
x}_j}( {\bxi} ) \ , \hspace{10mm} {J^e} = \det \left\{ \frac{\partial {\bx}
}{\partial {\bxi} } \right\} \ \mbox{for element $e$,} \hspace{10mm} Re
= \frac{{X} {V} \rho}{\mu} \ ,
\] 
$n$ is the number of nodes on each element, $\mu$ is viscosity, $\rho$
is density, $X$, $V$ are a length scale and a velocity scale
respectively, ${\bV}^{e \ linearisation}$ is the new second order
accurate linearisation (given in Subsection \ref{1}),
\[
{\bV}^{e \ linearisation} = 2 {\bV}^e \mid_t - {\bV}^e \mid_{t - \Delta
t},
\]
${\bV}^{e \ mesh}$ is the mesh velocity, $\epsilon$ is the penalty
parameter ($\epsilon = 10^{-6}$) and $g$ is the gravitational
acceleration. \footnotetext[2]{The $Q_2$--$P_1$ element pair was shown
to satisfy the L.B.B. condition in the context of rectangular elements.
It is important to note in this regard that a linear function mapped
from the master element using a $Q_2$ mapping will no longer be $P_1$
for non--rectangular elements.}
\renewcommand{\thefootnote}{\arabic{footnote}}

To recover the pressure on each element once the velocity solution has
been obtained,
\begin{eqnarray} \label{80}
p = - \frac{1}{\epsilon} \left[ \int_{\hat \Omega} \varphi_m \varphi_l
J^e d {\hat \Omega} \right]^{-1} \int_{\hat \Omega} \varphi_m G_{kjk}
J^e d {\hat \Omega} \ V_j^e \ \varphi_l.
\end{eqnarray}

\subsection{A Guideline for Artificial Reynolds Number Adjustment}

A straight forward Picard iteration process (in which the convective
term is linearised) may not be sufficient at higher Reynolds numbers
where the nonlinear term becomes more significant and the radius of
convergence is smaller. What one has to do under such circumstances is
implement a continuation technique. That is, lower the Reynolds number
to a level where convergence is attainable, then raise it
incrementally, obtaining progressively better solutions at each stage.
Choosing these various artificial Reynolds number levels can, however,
turn out to be somewhat problematic. What follows was devised as a
guideline in making this choice. Suppose the nonlinear and linear parts
of the equations are expressed in terms of two matrices, ${\bA}$ and
${\bB}$, respectively.  The equations to be solved are then
\[
A_{nkj}v_{k}v_{j} + B_{nj}v_{j} + c_n = 0.
\]
\begin{theorem} \label{8}
A sufficient condition for the iterative linearisation method to
converge is
\[
\mid \mid {\bD} \mid \mid_p < 1
\]
where 
\[
D_{jl} = (A_{nkj}v_{k}^{i-1} - B_{nj})^{-1}A_{nlm}v_{m}^{i-1}, 
\]
$\mid \mid . \mid \mid_p$ denotes the generalised $p$ norm and the
superscript $i$ denotes the number of the iteration from which a given
solution was obtained ({\sc Childs} and {\sc Reddy} \cite{me:3}).
\end{theorem}
{\sc Proof:} Let ${\bepsilon}^{i-1}$ denote the difference between the
solutions obtained from the $i$th and $(i-1)$th iterations. That is,
\[
{\bepsilon}^{i-1} = {\bv}^i - {\bv}^{i-1}
\] 
The system of equations solved at the $(i-1)$th Picard iterate was
accordingly
\begin{eqnarray} \label{6}
A_{nkj}v_{k}^{i-2}v_{j}^{i-1} + B_{nj}v_{j}^{i-1} + c_n &=& 0 \nonumber
\\ 
A_{nkj} (v_{k}^{i-1} - \epsilon_k^{i-2}) v_{j}^{i-1} + B_{nj}
v_{j}^{i-1} + c_n &=& 0.
\end{eqnarray}
The system of equations to be solved at the $i$th Picard iterate is
\begin{eqnarray} \label{7}
A_{nkj}v_{k}^{i-1}v_{j}^i + B_{nj}v_{j}^i + c_n &=& 0 \nonumber \\
A_{nkj}v_{k}^{i-1}(v_{j}^{i-1} + \epsilon_j^{i-1}) + B_{nj}(v_{j}^{i-1}
+ \epsilon_j^{i-1}) + c_n &=& 0.
\end{eqnarray}
Equating (\ref{6}) with (\ref{7}),
\begin{eqnarray*}
A_{nkj}v_{k}^{i-1} \epsilon_j^{i-1} + B_{nj} \epsilon_j^{i-1} &=& -
A_{nkj} \epsilon_k^{i-2} v_{j}^{i-1} \\
\epsilon_j^{i-1} &=& - (A_{nkj}v_{k}^{i-1} + B_{nj})^{-1} A_{nlm} \epsilon_l^{i-2} v_{m}^{i-1} \\
&=& D_{jl} \epsilon_l^{i-2} \\
\mid\mid {\bepsilon}^{i-1} \mid\mid_p &\leq& 
\mid\mid {\bD} \mid\mid_p \ \mid\mid {\bepsilon}^{i-2} 
\mid\mid_p. 
\end{eqnarray*}
A sufficient requirement for convergence is therefore
\begin{eqnarray*}
\mid \mid {\bepsilon}^{i-1} \mid \mid_p \ < \ \mid \mid
{\bepsilon}^{i-2} \mid \mid_p &\Leftrightarrow& \mid
\mid {\bD} \mid \mid_p < 1
\end{eqnarray*}

\begin{corollary}
A sufficient condition for the iterative linearisation method to
converge is
\[
\max_l \ \Sigma_{j=1}^{n} \mid (A_{nkj}v_{k}^{i-1} -
B_{nj})^{-1}A_{nlm}v_{m}^{i-1} \mid < 1
\] 
(the maximum absolute column sum $<$ 1), where the superscript $i$
denotes the number of the iteration from which that particular solution
was obtained.
\end{corollary}
{\sc Proof:} In the particular instance of the $1$ vector norm,
\[
\mid \mid {\bD} \mid \mid_1 = \max_l \ \Sigma_{j=1}^{n} \mid
(A_{nkj}v_{k}^{i-1} - B_{nj})^{-1}A_{nlm}v_{m}^{i-1} \mid 
\]
(The natural matrix norm induced by the 1 vector norm is the maximum
absolute column sum.)

{\sc Remark:} This guideline serves only to examine local convergence,
and one still has to hunt for a suitable Reynolds number by trial and
error. It does, however, is provide an exact quantitative measure of
this local convergence and, depending on the solver used, can improve
efficiency.

\section{A Predictor--Corrector or ``Linearised'' Approach} \label{5}

The combined free surface--fluid--rigid body problem is highly
nonlinear. An important decision had to be taken on whether to solve
the respective sub--problems simultaneously, using a Newton--Raphson
method, or separately using a predictor--corrector (iterative)
approach. The latter approach may be justified in terms of a
Picard-type linearisation process (similar to that discussed in Section
\ref{1} when dealing with the convective term).

The advantages of a predictor--corrector approach over solving
simultaneously (the approach taken by many) are expected to be a larger
radius of convergence and hence a more robust scheme, substantially
less code and easy, separate testing of sub--problems.

The disadvantages of a predictor--corrector approach over solving
simultaneously are expected to be linear, as opposed to quadratic,
rates of convergence. The expected slow rate of convergence could
possibly be overcome to a degree by taking good initial guesses based
on previous solutions, using linear extrapolation for example.

\subsection{The Resulting Algorithm}

\begin{figure}[H]
\begin{center} \leavevmode
\mbox{\epsfbox{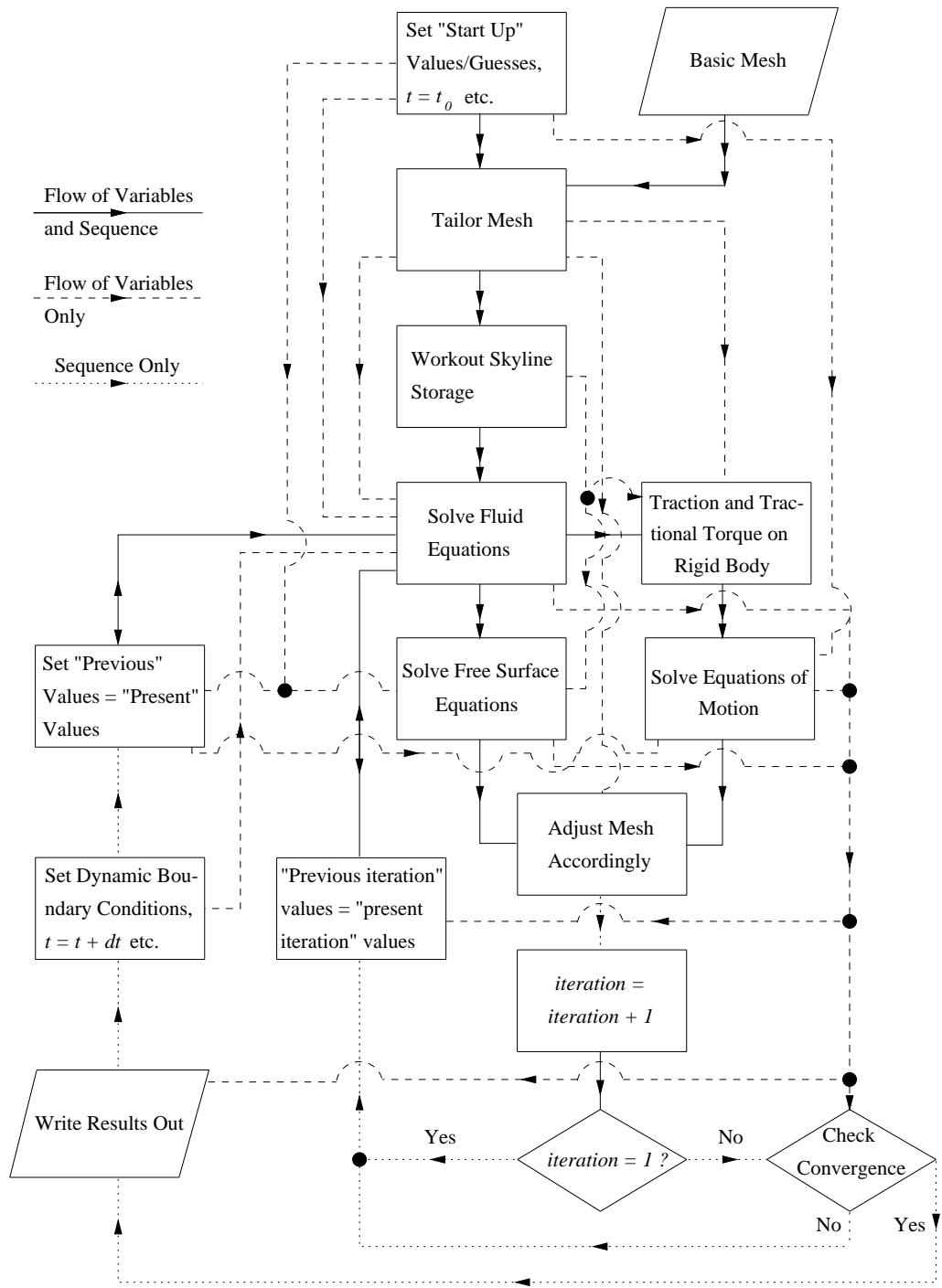}}
\end{center}
\caption{Combined Structure--Flow Chart Outlining the
Predictor--Corrector Algorithm for the Fluid--Rigid--Body--Free Surface
Problem.} \label{109}
\end{figure}

Figure \ref{109} outlines the algorithm which results from the use of
such a linearised approach. The solution values of the fluid
sub--problem are used in solving the free surface sub--problem,
working out the traction on the rigid body and consequently in
solving the rigid body sub--problem. The mesh is then adjusted
accordingly and, if not the first iteration, the convergence is
examined. If only the first iteration, alternatively, if the degree
of convergence is found to be unsatisfactory, the process is
repeated. Once this iteration process has converged, a new time step
is commenced.

The O(${\Delta t}^4$) Runge--Kutta--Fehlberg and Adams--Moulton
approximations used in the rigid body sub--problem are possibly
excessive in comparison to the  O(${\Delta t}$) finite difference
schemes implemented in the fluid and free surface problems.

\section{Automatic Mesh Generation About Included Bodies of Any Shape}
\label{9}

The finite element mesh was automatically generated and adjusted about
the included rigid body in what is possibly a slightly novel fashion. A
small region of mesh immediately adjacent to the included rigid body
was repeatedly remapped to cope with the changing orientation, the
remainder was squashed/stretched according to the translation.

\subsubsection*{Local Distortion About an Included Rigid Body}

\begin{figure}[H]
\begin{center} \leavevmode
\mbox{\epsfbox{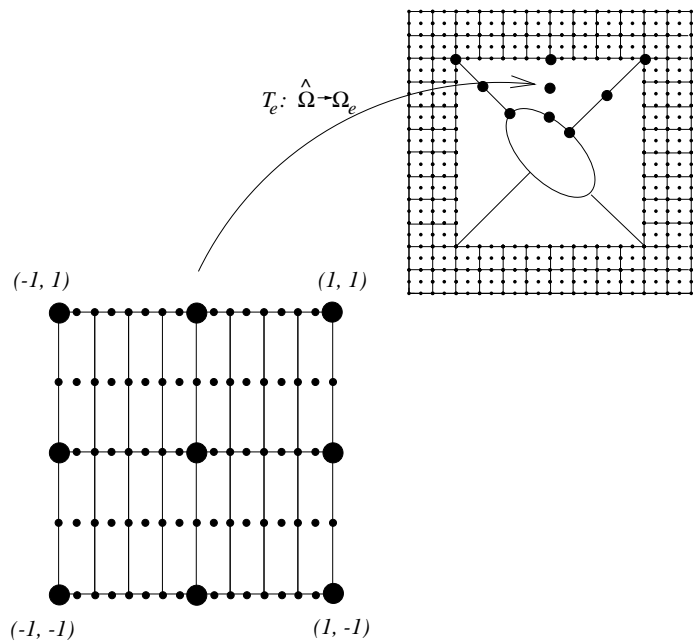}}
\end{center}
\caption{The Local Distortion is Obtained by Mapping Square Chunks of
Rectangular Mesh Using Finite Element Mappings.} \label{3}
\end{figure}

The local distortion about included rigid bodies is obtained by mapping
four square chunks of rectangular mesh to the four wedge--shaped
domains depicted in Figure \ref{3} using finite element mappings.  A
square region of mesh centered on, and including the rigid body, is
first removed. Each of the depicted wedge--shaped regions is then
demarcated by as many points as there are nodes in an element i.e. each
wedge shaped--region is set up as a massive element.

{\bf Demarcation:} Demarcation is accomplished by first locating the
intersection of the lines which bisect corners and edges of the square
frame, with the surface of the rigid body using Newton's method. The
remainder of this demarcation process needs no explanation.
{\bf Mapping:} Chunks of uniform mesh, which have identical extremities
to those of the master element, are then mapped into the
newly--demarcated, wedge--shaped regions using finite element mappings.
The devised method maps chunks of uniform mesh into wedge--shaped
regions surrounding the included rigid body in exactly the same manner
as points in the master element domain are, in theory, mapped into
individual mesh elements. In the finite element mapping,
\begin{eqnarray*}
T_e: {\hat \Omega} &\rightarrow& \Omega_e \\
{\bx}({\bxi}) &=& \sum_{i=1}^{\mbox{\scriptsize \it nNode}} {\bx}_i \phi_i({\bxi}),
\end{eqnarray*}
symbols which previously denoted the master element, the master element
variables, the corresponding position in the domain, the position of
node $i$ and element $e$ now denote the following:  $\hat \Omega$ --
the chunk of uniform mesh, $\bxi$ -- the position of a constituent node
in the chunk, ${\bx}$ -- the corresponding position in the wedge,
${\bx}_i$ -- the position of the $i$th point demarcating the wedge and
$\Omega_e$ -- the wedge.

{\bf Interface Adjustment:} Further fine adjustment of nodes intended
to delineate the surface of the rigid body is accomplished by moving
them along a line between node and centre, to the rigid body surface
using Newton's method, thereby accomodating additional shape
complexity.
\begin{figure}[H]
\begin{center} \leavevmode
\mbox{\epsfbox{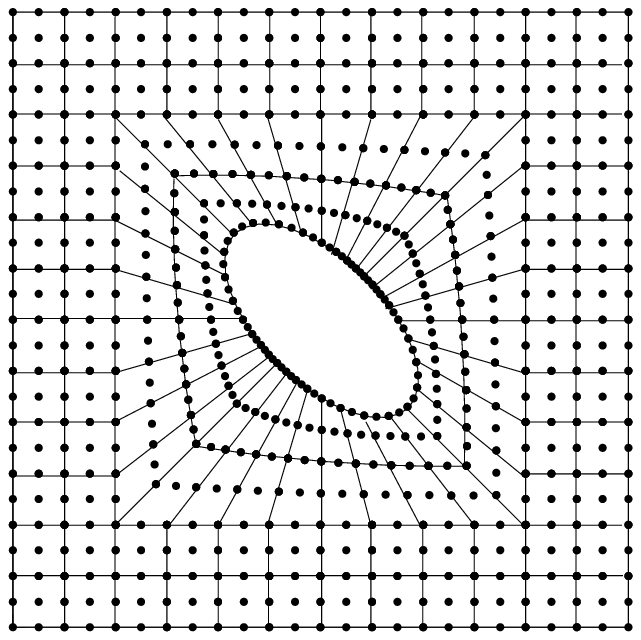}}
\end{center}
\caption{Automatic Mesh Generation about a Rigid Body which is
Simultaneously Rotating and Translating.}
\end{figure}
The method described was found to be remarkably practical,
simple and effective with maximum angles never exceeding
$\frac{\pi}{4}$ radians at element vertices within the mesh.

\subsubsection*{Translational Mesh Deformation}

The mesh outside the ``box'' (the box containing the 4 ``wedges''
enclosing the rigid body) is squashed/stretched according to the
requirements of the translation (the nodes are translated in the same
direction by a factor inversely proportional to their distance from the
box).
\begin{figure}[H]
\begin{center} \leavevmode
\mbox{\epsfbox{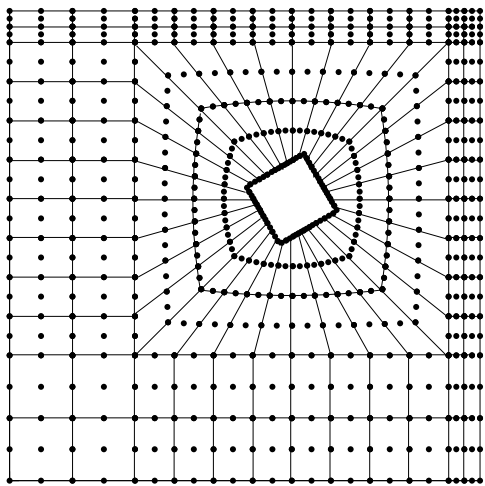}}
\end{center}
\caption{Automatic Mesh Generation and Adjustment about a Diamond which
is Simultaneously Rotating and Translating.} \label{169}
\end{figure}

\section{Some Test Examples} \label{4}

The final model was tested in the context of a driven cavity flow, a
driven cavity flow with various, included rigid bodies, a die--swell
problem and a Stokes, second--order wave. The idea was to use as simple
as possible flows at extremely low Reynolds numbers, so as to generate
the smoothest possible flows in which any resulting rigid body motion
could be expected to be highly predictable, alternatively examples for
which there exist well established results. 

\subsection{Example 1: Driven Cavity Flow} \label{2} 

A variety of bench-mark tests for viscous, incompressible flows are
cited in the literature. The driven cavity flows of {\sc Carey} and
{\sc Krishnan} \cite{c:1} and {\sc Simo} and {\sc Armero} \cite{s:1}
are two such examples. The so--called ``leaky lid'' driven cavity flow
of {\sc Simo et. al.} was found to be the more sensitive of the two
when testing the performance of schemes approximating the
Navier--Stokes equations. In addition, {\sc Simo et. al.} tested the
scheme for the Navier-Stokes equations rigorously using the leaky lid
problem.

These results could be used as a comparison in testing the performance
of the completely general referential description derived in {\sc
Childs} \cite{me:1} as well as to compare the performance of the
standard, penalty method (used here) with that of the iterative,
augmented Lagrangian approach (used by {\sc Simo et.  al.}). The main
objective of these tests was nonetheless to generate and study as
smooth as possible a flow into which a somewhat inconsequential
(flow--wise inconsequential) rigid body could later be introduced (see
the forthcoming example).

{\sc Remark:} Both the above mentioned flows are steady in the context
of the traditional Eulerian description. They are, however,
``unsteady'' in the context of a deforming mesh, a factor enhancing
their potential as tests.

\begin{figure}[H]
\begin{center} \leavevmode
\mbox{\epsfbox{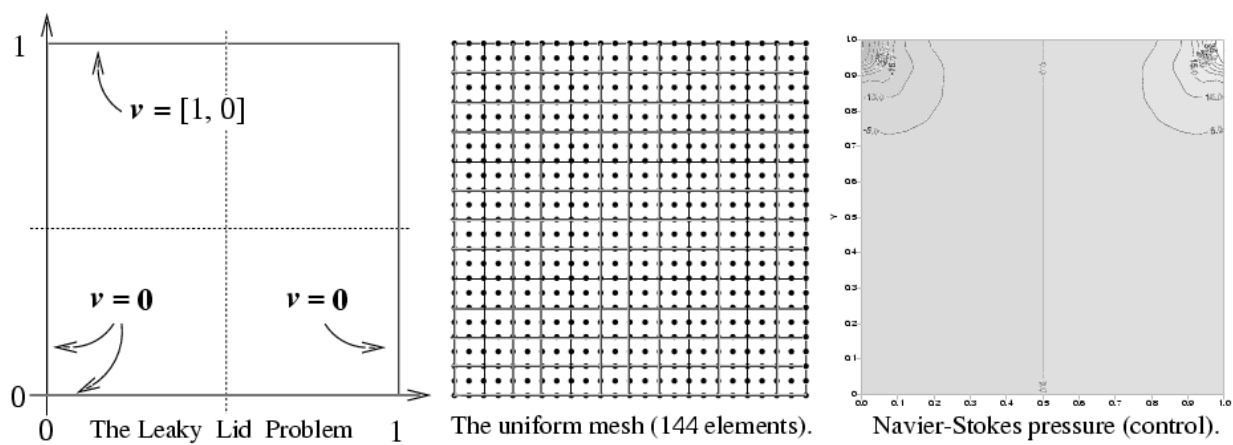}}
\end{center}
\caption{The Problem, the Mesh and the Pressures Obtained Using the
Conventional Eulerian Equations.} \label{82}
\end{figure}

The problem is that of a square, two--dimensional ``pot'' whose lid is
moved across the top at a rate  equal to its diameter for a Reynolds
number of unity.  The boundary conditions are accordingly ``no slip''
on container walls and a horizontal flow of unity across the top (as
depicted in Figure \ref{82}). The problem also bears a certain
resemblance to an idealised pothole in a river bed. A mesh consisting
of one hundred and forty four elements was deformed at differing rates
(also depicted in Figure \ref{82}).

\subsubsection*{Results}

\begin{figure}[H]
\begin{center} \leavevmode
\mbox{\epsfbox{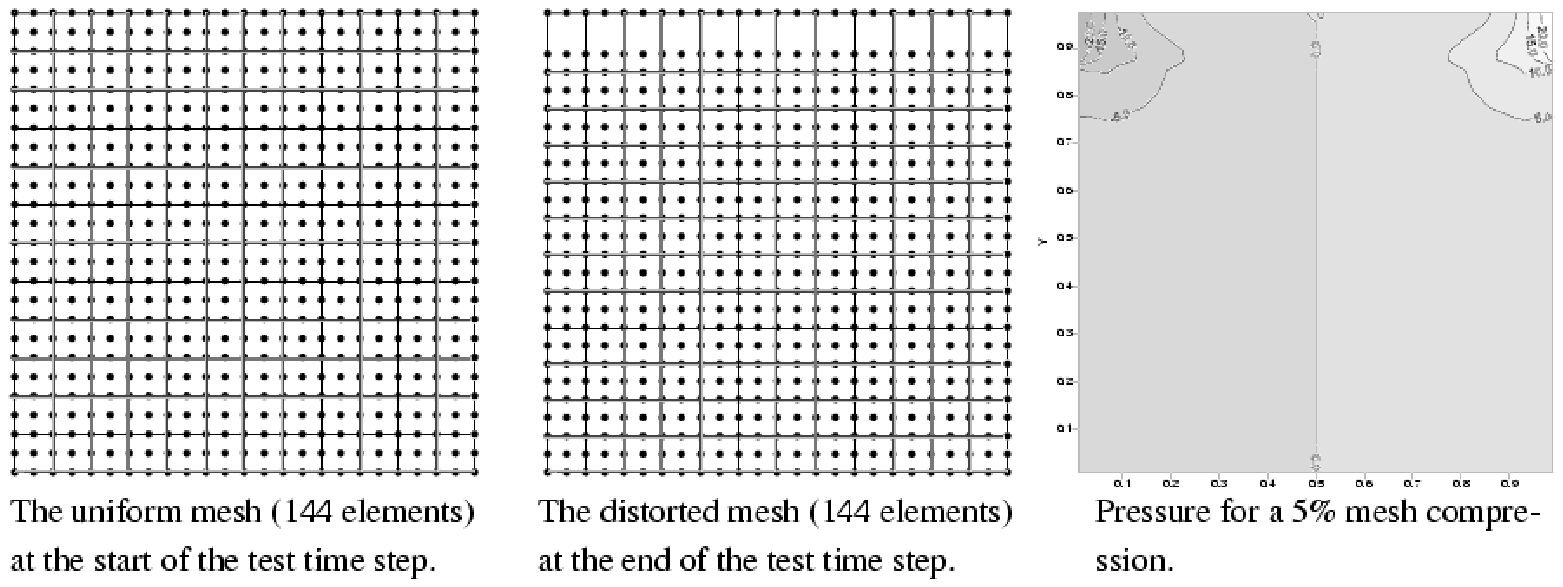}}
\end{center}
\caption{Pressures Obtained Using the Completely General Referential
Equation.} \label{76}
\end{figure}

\begin{figure}[H]
\begin{center} \leavevmode
\mbox{\epsfbox{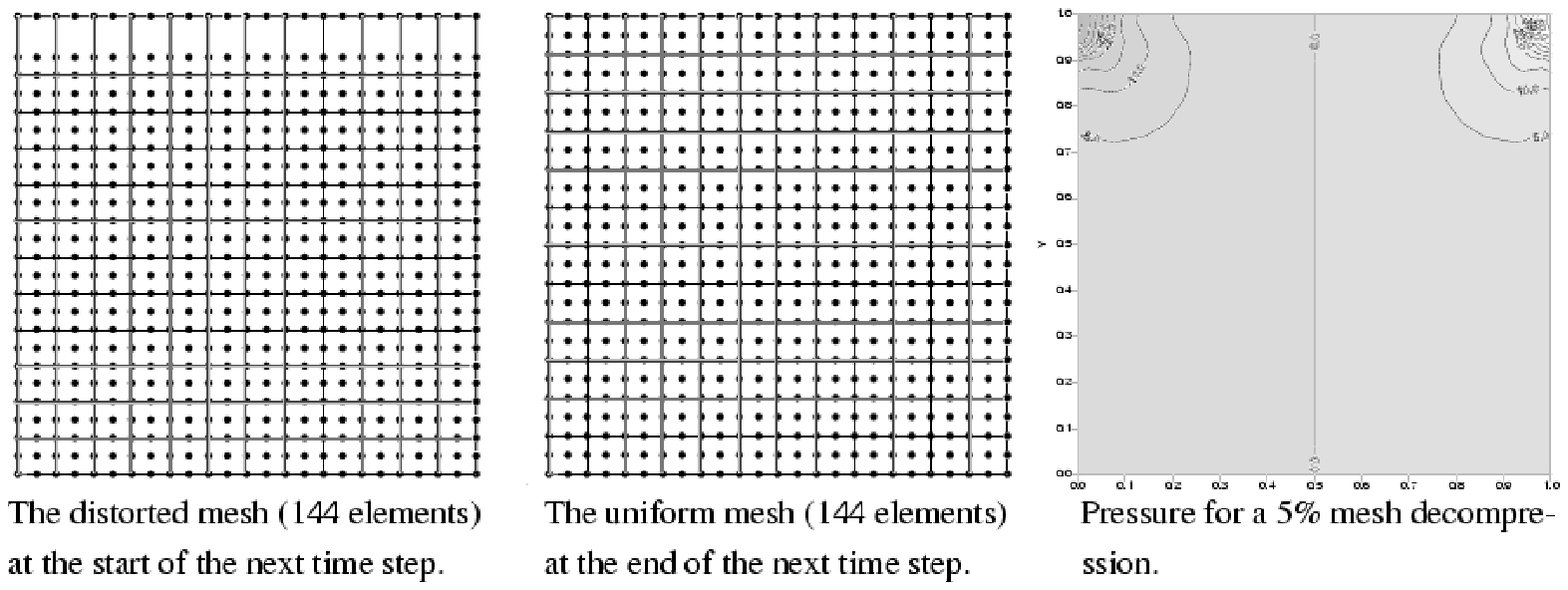}}
\end{center}
\caption{Pressures Obtained Using the Completely General Referential
Equation.} \label{83}
\end{figure}

\begin{figure}[H]
\begin{center} \leavevmode
\mbox{\epsfbox{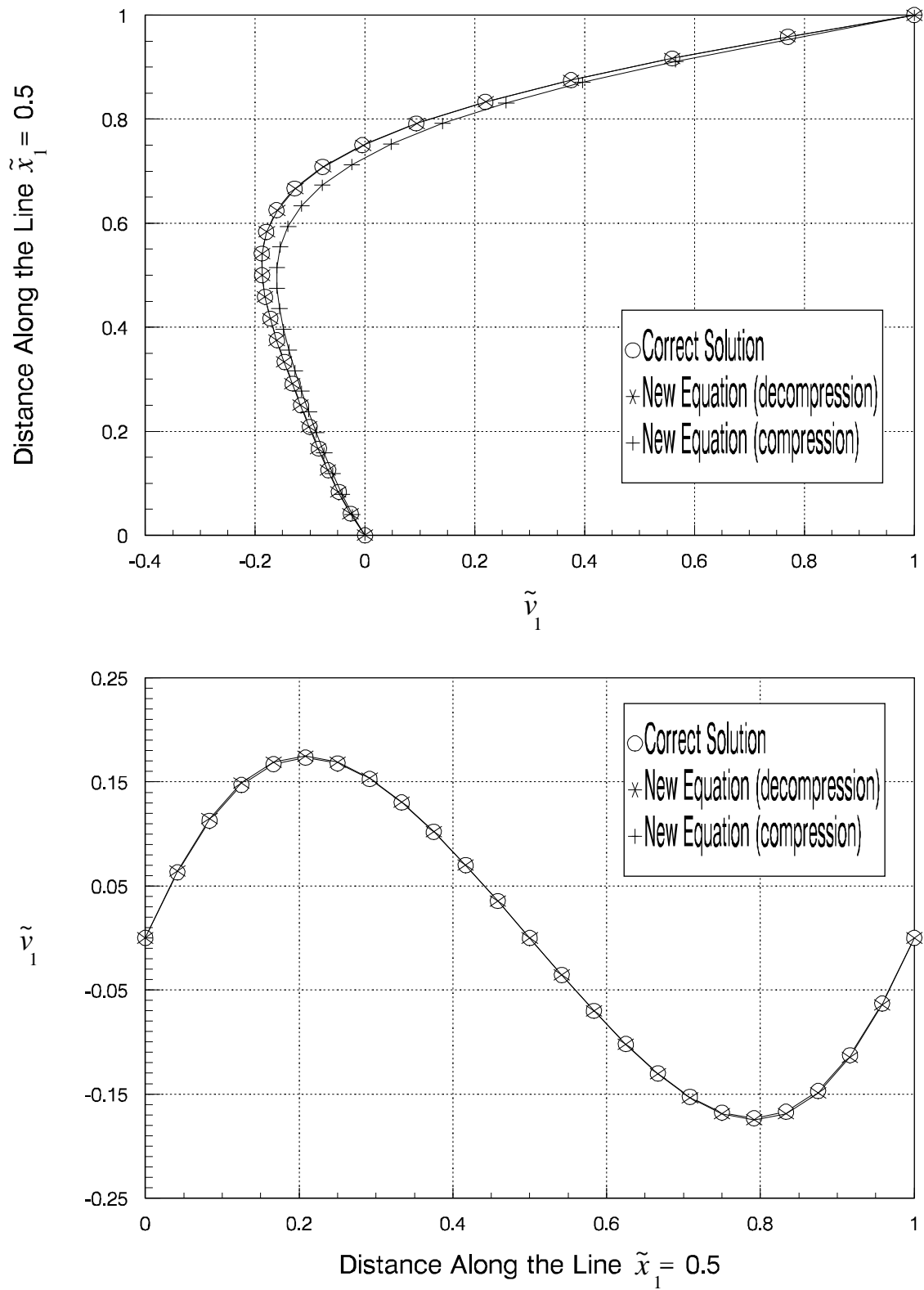}}
\end{center}
\caption{In this test part of the mesh was successively compressed
and decompressed by 5 \% over two time steps of length 0.05.} \label{84}
\end{figure}

The results in Figures \ref{76}, \ref{83} and \ref{84} obtained while
deforming the 144 element mesh depicted, are in agreement with those
obtained using the analogous Navier--Stokes algorithm on a fixed
uniform mesh (the control results in Figure \ref{82} and the relevant
velocity profiles in Figure \ref{84} on page \pageref{84}). They are
also in agreement with the results of {\sc Simo} and {\sc Armero}
\cite{s:1}, obtained using the Navier--Stokes equations (the
conventional Eulerian description) on a fixed, uniform mesh consisting
of 3200 triangular elements.
The fixed, uniform mesh, with which the control results were obtained,
consisted of 144 9--noded, quadrilateral elements. In the first test
all, but the 25 upper boundary nodes, were compressed into the lower 95
\% of the problem domain after which the mesh was restored to a uniform
state.
The mesh was successively compressed and decompressed over two time
steps, each of length 0.05. Mesh velocities were therefore of the order
of 10 times greater than the flow velocities being modelled.

\newpage
\begin{center} {\large Velocity Profiles} \end{center}
\vspace{10mm}
\newpage

\begin{figure}[H]
\begin{center} \leavevmode
\mbox{\epsfbox{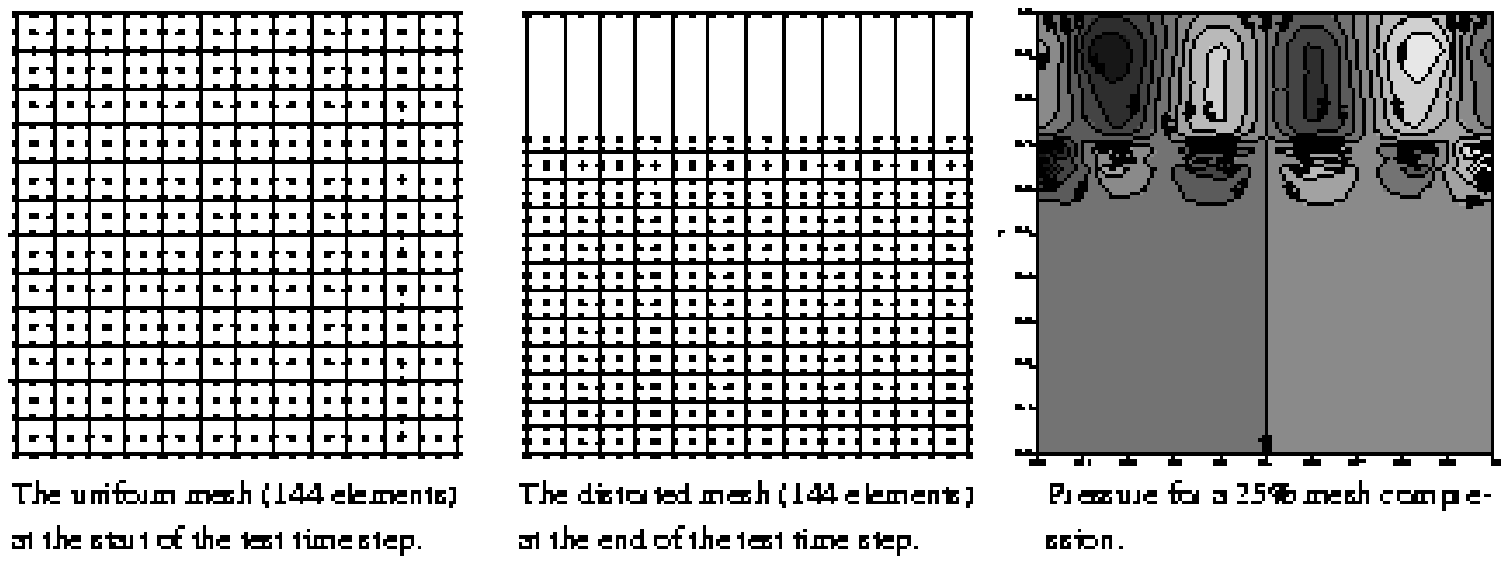}}
\end{center}
\caption{Pressures Obtained Using the Completely General Referential
Equation.} \label{77}
\end{figure}

\begin{figure}[H]
\begin{center} \leavevmode
\mbox{\epsfbox{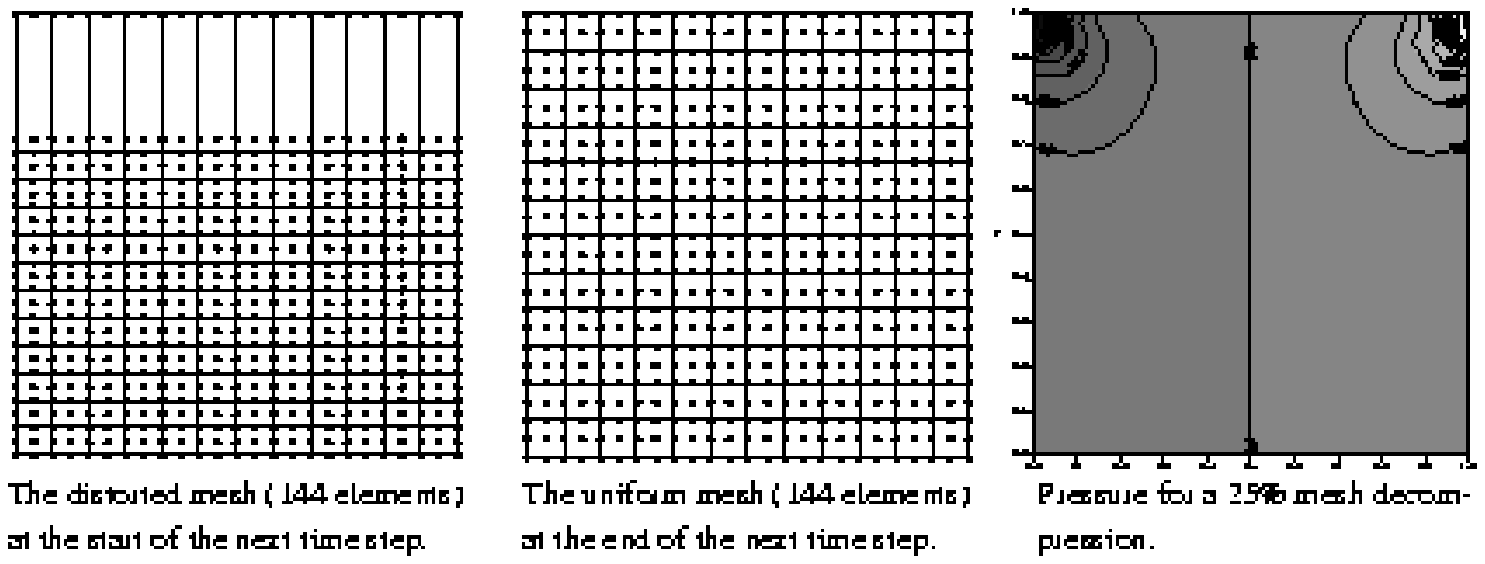}}
\end{center}
\caption{Pressures Obtained Using the Completely General Referential
Equation.} \label{85}
\end{figure}

\begin{figure}[H]
\begin{center} \leavevmode
\mbox{\epsfbox{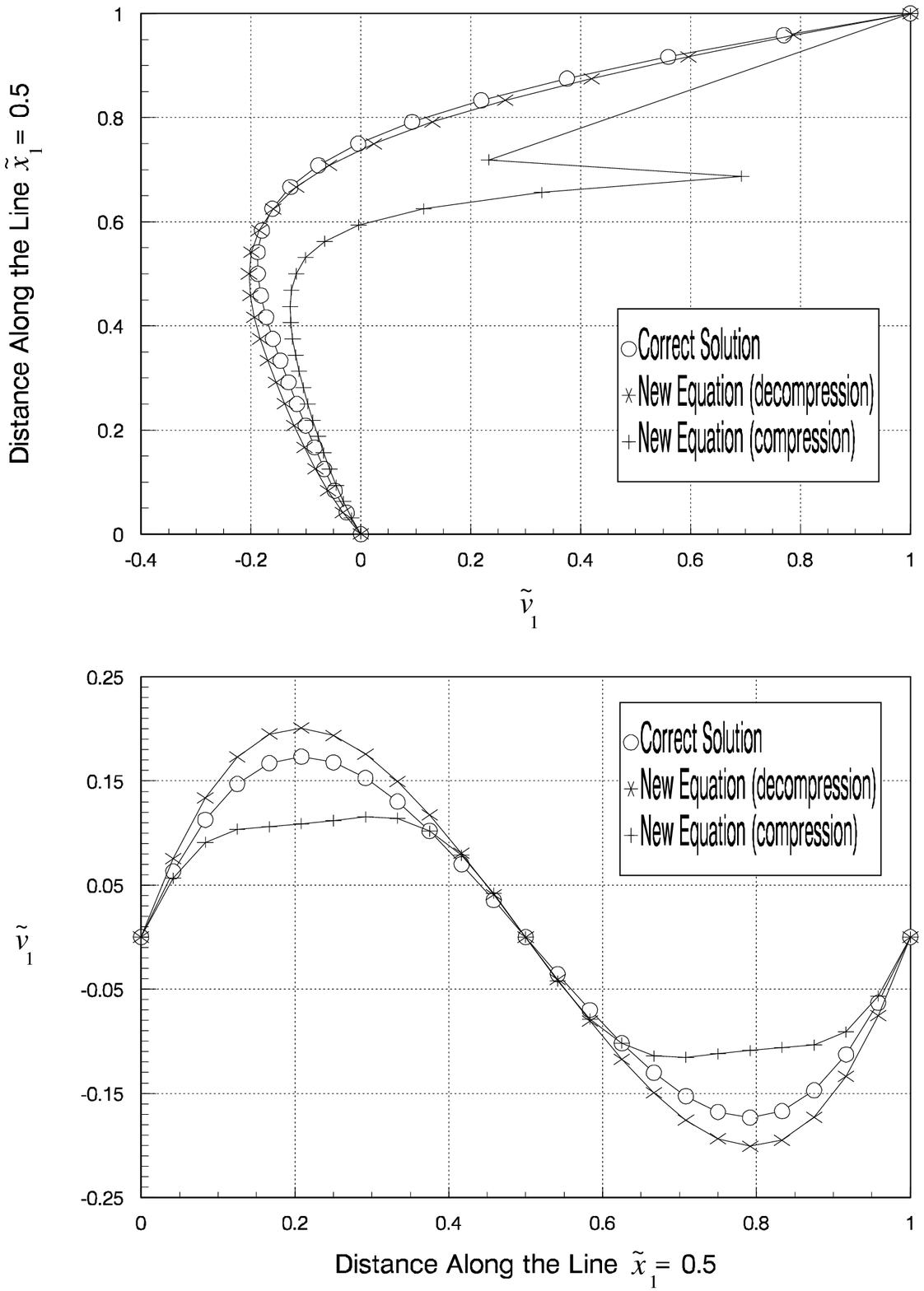}}
\end{center}
\caption{In this test part of the mesh was successively compressed
and decompressed by 25 \% over two time steps of length 0.05.} \label{86}
\end{figure}

In the second test (Figs. \ref{77}, \ref{85} and \ref{86}) all but 25
boundary nodes were compressed into the lower three quarters of the
domain after which the mesh was restored to a uniform state. 
In this example the mesh was successively compressed
and decompressed by 25 \% over two time steps of length 0.05. Mesh
velocities were therefore of the order of 50 times greater than the
flow velocities being modelled.
The error is attributed to a combination of a bad mesh (mostly) and
relatively high off--scale mesh velocities. It should be remembered
that a bad mesh is a bad mesh nonetheless.

\newpage
\begin{center} {\large Velocity Profiles} \end{center}
\vspace{10mm}

{\sc Remark:} A further fact illustrated by the results of this test
is that, of the pressure and velocity solutions, the pressure
solution is the more sensitive of the two. This is to be expected
since the pressure is recovered from what is fundamentally a
difference in something like the tenth decimal place of the velocity
solution (see equation (\ref{80})) depending on the penalty
parameter.

\subsection{Example 2: ``Pebble in a Pothole''}

In this example rigid bodies of varying mass and moments of inertia
were released from rest in a flow dictated by the same boundary
conditions as the driven cavity flow of the previous example. One
would expect a ``die bead'' (a small rigid body of neutral bouyancy)
to move in tandem with the fluid soon after its release from rest.
One might also expect a clockwise spinning to be induced by
concentrating the mass closer to the centre i.e. lowering the moment
of inertia.

Various rigid bodies were introduced to the cavity driven flow
problem described in Subsection \ref{2}, in the absence of a body
force. Although the algorithm allows for the inclusion of a rigid
body of any shape, one which could be expected to behave fairly
predictably was selected for the purposes of the test. The results
presented shortly involve the elliptical rod--shaped body
\[
\frac{x_1^2}{2^2} + \frac{x_2^2}{1^2} = 0.025^2,
\]
whose major axis is $0.1$.

\subsubsection*{Results}

\begin{figure}[H]
\begin{center} \leavevmode
\mbox{\epsfbox{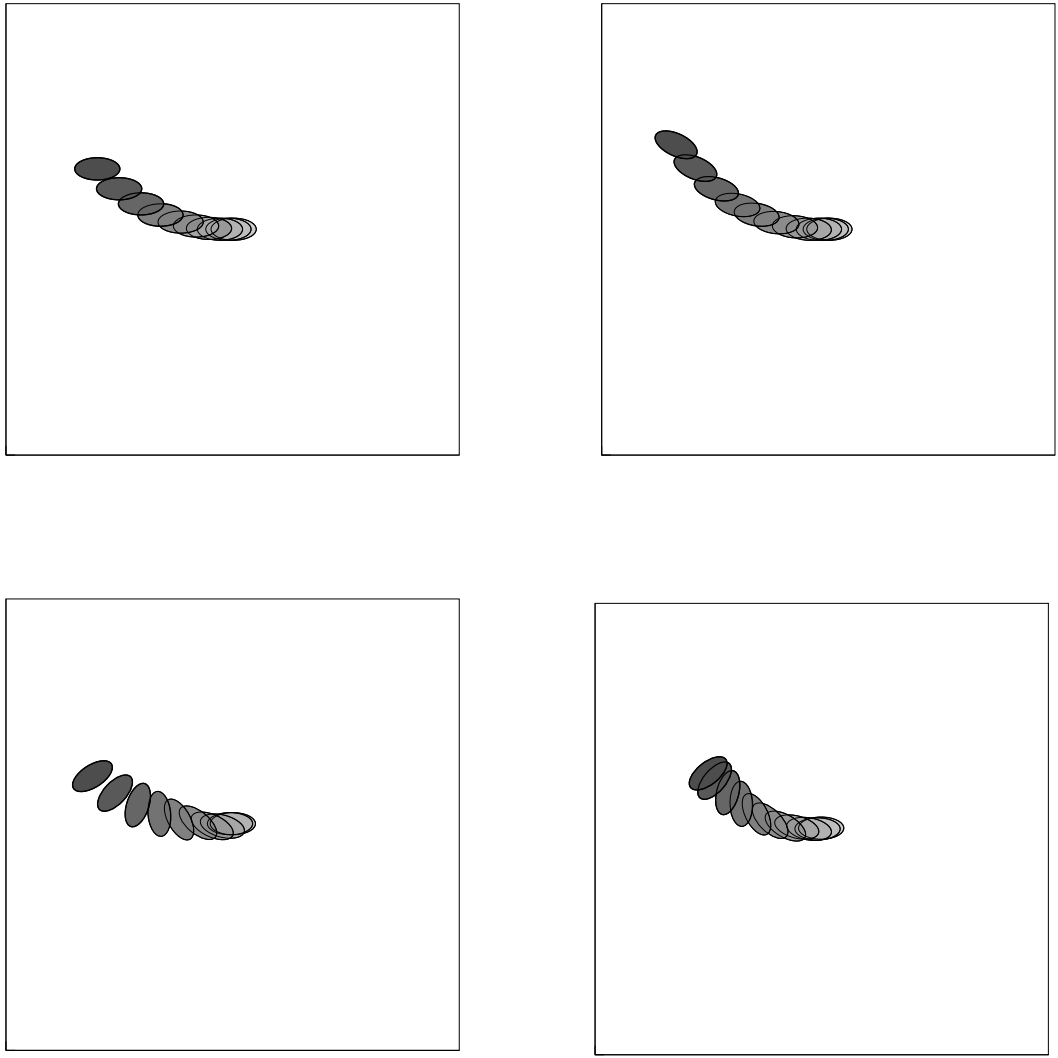}}
\end{center}
\caption{The trajectories of various included rigid bodies released
from rest at the centre of the driven cavity flow described in
Subsection \ref{2}. {\sc Top Left:} $Re = 0.025$, ${\bar m} = 251.3$,
${\bar J}_{33} = 314.2$ and $t = 3.6$ secs. {\sc Top Right:} $Re =
0.025$, ${\bar m} = 251.3$, ${\bar J}_{33} = 1.0$ and $t = 4.0$ secs.
{\sc Bottom Left:} $Re = 0.025$, ${\bar m} = 251.3$, ${\bar J}_{33} =
0.1$ and $t = 3.6$ secs. {\sc Bottom Right:} $Re = 1$, ${\bar m} = 1$,
moment of inertia (scaled) $= 0.1 $ and $t = 2.0$ secs.} \label{108}
\end{figure}

The results which follow were in agreement with the notion that the die
bead would move in tandem with the fluid soon after its release from
rest. In the succesive trajectories in Figure \ref{108} the mass was
successively concentrated closer to the centre (a lower moment of
inertia was used). A clockwise spin was then induced.

The last trajectory in Figure \ref{108} was obtained using a Reynolds
number of unity in order to generate a smooth, predictable flow as
close as possible to the driven cavity flow of Subsection \ref{108}.
Although the scaling is not immediately reminiscent of any real life
problem, the example serves to further verify the potential of the
methods described qualitatively.

The results for the included rigid body tests were extremely
encouraging. This is especially so when it is considered that the
Reynolds numbers implied by small, included rigid bodies are low and
the model being developed is eventually intended to elucidate problems
of a sedimentological nature.

\subsection{Example 3: Die Swell Problems}

\begin{figure}[H]
\begin{center} \leavevmode
\mbox{\epsfbox{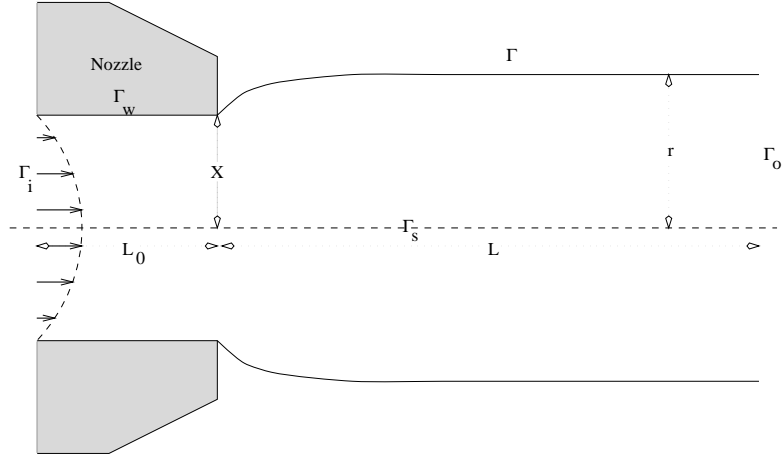}}
\end{center}
\caption{The Die Swell Problem} \label{90} 
\end{figure}

The axis--symmetric die swell (or fluid jet) problem is a free surface
problem well documented in the literature. The exact details of the
problem description always differ to varying degrees, however, the
central theme basic to all involves the extrusion of a fluid with
initial parabolic flow profile from the end of a short nozzle. The
various boundaries in terms of which this particular version is
specified are depicted in Figure \ref{90}.
\begin{tabular}{l r c l}
At the free surface, ${\Gamma}$: & & & \\
& $h \mid_{x_1 = 0}$ &$=$& $X$ \\
& & & \\
& $\left. \displaystyle \frac{d h}{d x_1} \right|_{x_1 = L}$ &$=$& $0$
\\
& & & \\
At the nozzle wall, ${\Gamma}_w$: & & & \\
& ${\bv}$ &$=$& $0$ \\
& & & \\
At the inlet, ${\Gamma}_i$: & & & \\
& $v_1$ &$=$& $ \displaystyle \frac{3}{2} V \left( 1 -
\displaystyle \frac{x_2^2}{X^2} \right)$ \\
& & & \\
& &$=$& $\displaystyle \frac{3}{2} \displaystyle \frac{\mu}{\rho X}
\mbox{\em Re} \left( 1 - \displaystyle \frac{x_2^2}{X^2} \right)$ \\
& & & \\
& $v_2$ &$=$& $0$ \\
& & & \\
At the outlet, ${\Gamma}_o$: & & & \\
& ${\bsigma}{\bn} \cdot {\bn}$ &$=$& $0 \hspace{10mm} (= \ {\bar
{\bP}}{\bar {\bN}} \cdot {\bar {\bN}})$ \\
& & & \\
& ${\bv} \cdot {\bt}$ &$=$& $0 \hspace{10mm} (= \ {\bar {\bv}} \cdot
{\bar {\bT}})$ \\
& & & \\
At the symmetry axis, ${\Gamma}_s$: & & & \\
& ${\bsigma}{\bn} \cdot {\bt}$ &$=$& $0 \hspace{10mm} (= \ {\bar
{\bP}}{\bar {\bN}} \cdot {\bar {\bT}})$ \\
& & & \\
& ${\bv} \cdot {\bn}$ &$=$& $0 \hspace{10mm} (= \ {\bar {\bv}} \cdot
{\bar {\bN}})$
\end{tabular}

{\bf The Symmetry Axis, ${\Gamma}_s$}: Simultaneous implementation of
both ${\Gamma}_s$ boundary conditions amounts to leaving the term
\[
\int_{{\bar \Gamma}_s} {\bar {\bw}} \cdot {\bar {\bP}} {\bar {\bN}}
\ d {{\bar \Gamma}_s}
\]
out of the fluid equations and specifying ${\bar {\bv}} \cdot {\bar
{\bN}}$, the component of velocity normal to the boundary. Because the
boundary condition is Dirichlet in the normal direction, variational
formulation is consequently not required in that direction. This allows
${\bar {\bw}}$ in
\begin{eqnarray} \label{81}
\int_{{\bar \Gamma}_s} {\bar {\bw}} \cdot {\bar {\bP}} {\bar {\bN}}
\ d {{\bar \Gamma}_s} = \int_{{\bar \Gamma}_s} ({\bar {\bP}}
{\bar {\bN}} \cdot {\bar {\bN}} ) ({\bar {\bw}} \cdot {\bar {\bN}} )
\ d {{\bar \Gamma}_s} + \int_{{\bar \Gamma}_s} ({\bar {\bP}}
{\bar {\bN}} \cdot {\bar {\bT}} ) ({\bar {\bw}} \cdot {\bar {\bT}} )
\ d {{\bar \Gamma}_s}
\end{eqnarray}
to be chosen arbitrarily, but in such a way that ${\bar {\bw}} \cdot
{\bar {\bN}} = 0$.

{\bf The Outlet, ${\Gamma}_o$}: Implementing both ${\Gamma}_o$ boundary
conditions simultaneously amounts to omitting the term
\[
\int_{{\Gamma}_o} {\bar {\bw}} \cdot {\bar {\bP}} {\bar {\bN}}
\ d {{\Gamma}_o}
\]
and specifying ${\bar {\bv}} \cdot {\bar {\bT}}$. This implementation
can be justified in a likewise fashion to that given for the boundary
condition at the symmetry axis.

\subsubsection*{Results}

Examples of die swell problems in which $X = 1$, $L_o = 3.5$ and $L =
20$ for various kinematic viscosities and/or velocity scales were
examined using a 150 (50 $\times$ 3) element mesh. The predicted die
swell ratios agree well with those obtained by other, more specific
methods.

\begin{figure}[H]
\begin{center} \leavevmode
\mbox{\epsfbox{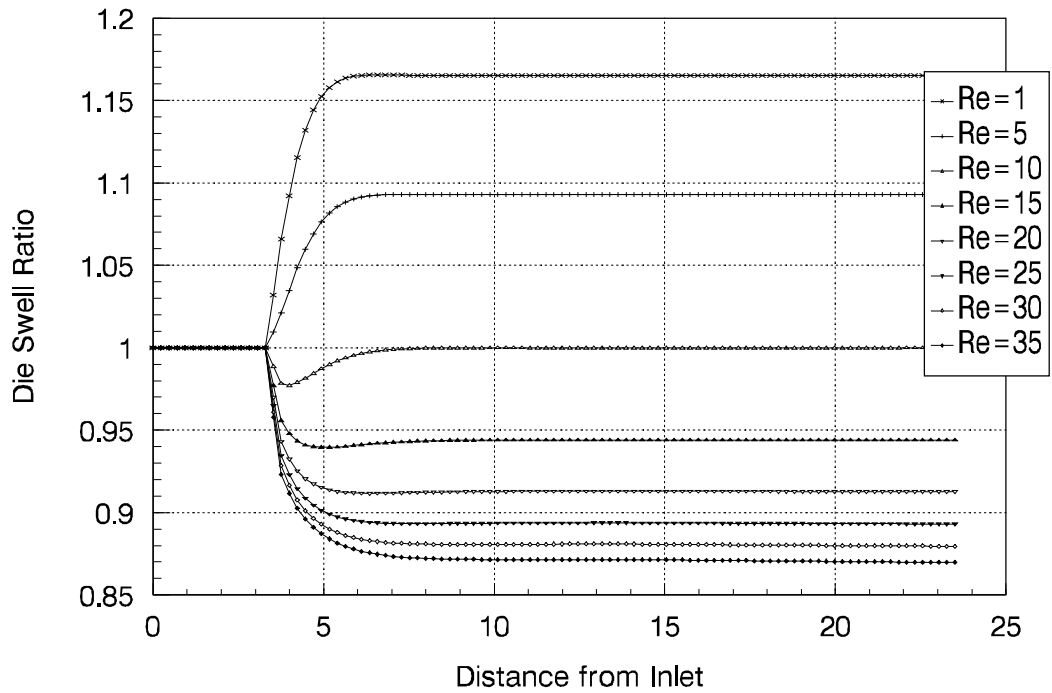}}
\end{center}
\caption{Die swell ratios predicted for various Reynolds numbers using
an inlet velocity profile of ${\bar v}_1 = \frac{Re}{1} \frac{3}{2}
(1-{{\bar x}_2^2})$ and the methods described.} \label{131}
\end{figure}

\begin{figure}[H]
\begin{center} \leavevmode
\mbox{\epsfbox{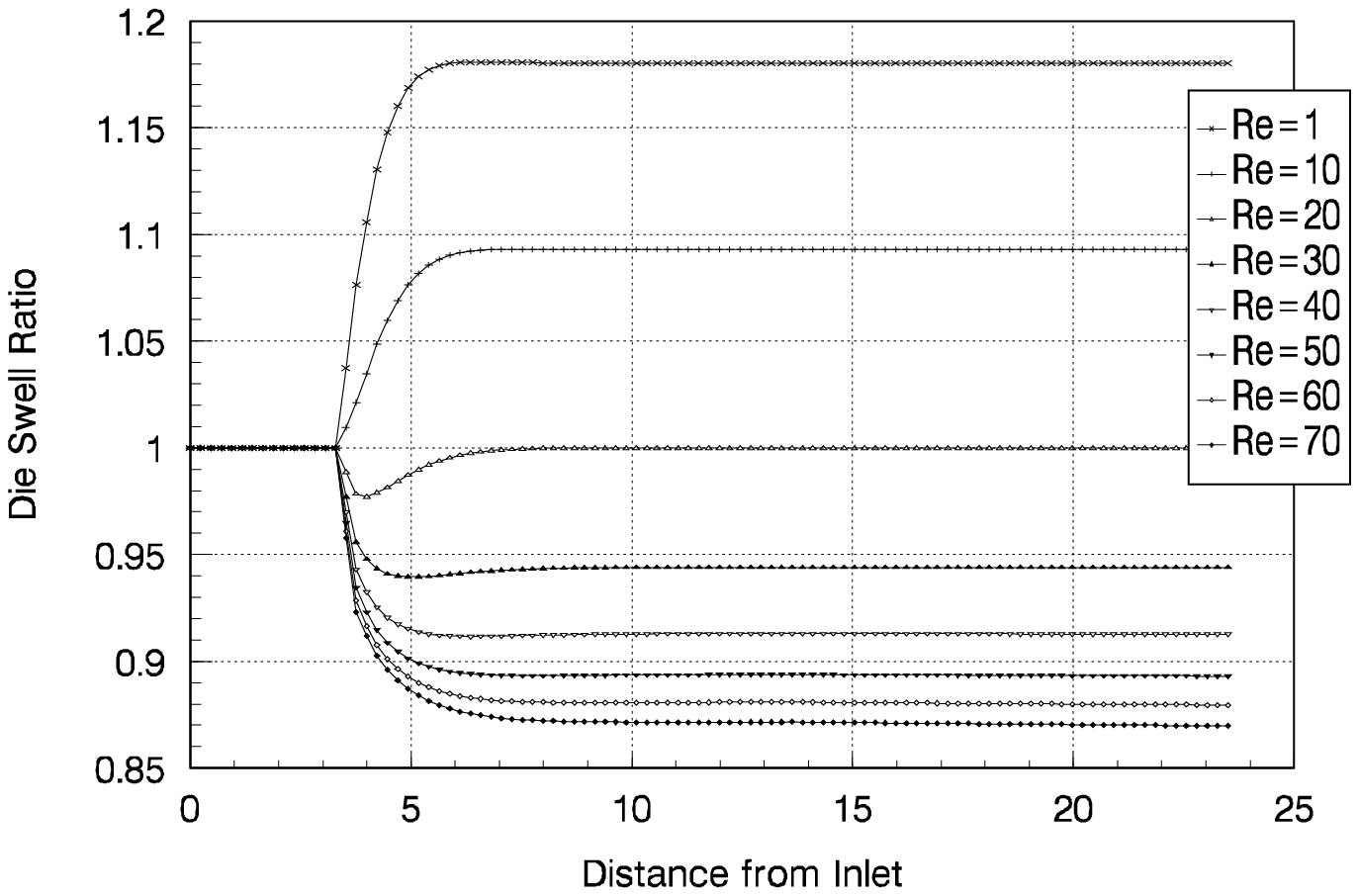}}
\end{center}
\caption{Die swell ratios predicted for various Reynolds numbers using
an inlet velocity profile of ${\bar v}_1 = \frac{Re}{10} \frac{3}{2}
(1-{{\bar x}_2^2})$ and the methods described.} \label{132}
\end{figure}

\begin{figure}[H]
\begin{center} \leavevmode
\mbox{\epsfbox{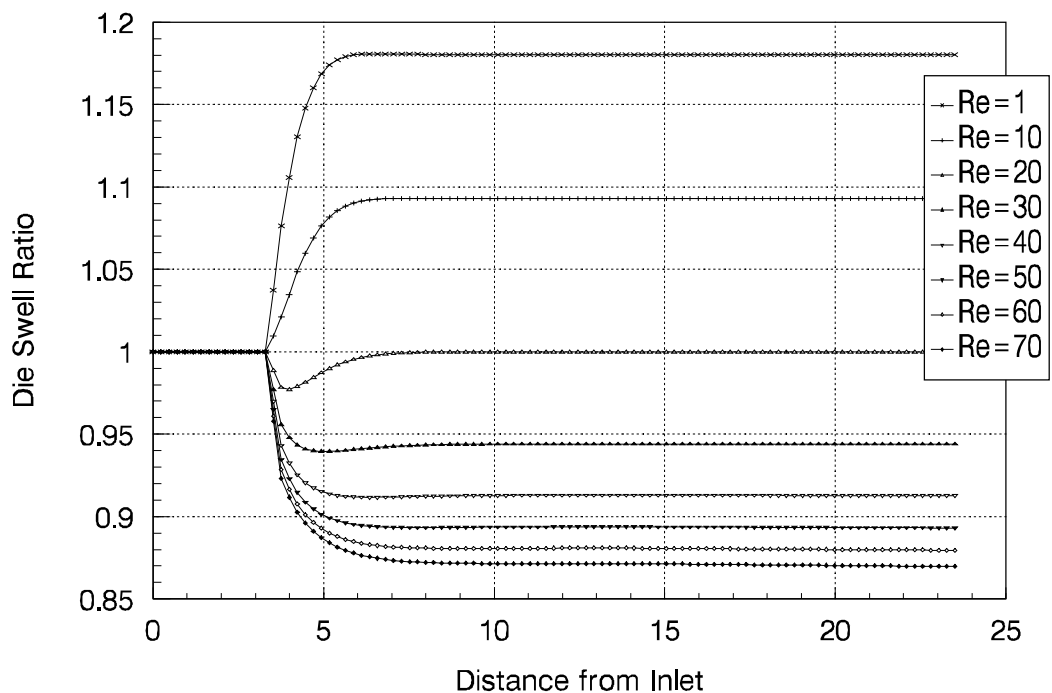}}
\end{center}
\caption{Die swell ratios predicted for various Reynolds numbers using
an inlet velocity profile of ${\bar v}_1 = \frac{Re}{20} \frac{3}{2}
(1-{{\bar x}_2^2})$ and the methods described.} \label{64}
\end{figure}

Compare \footnotemark[1] the result of {\sc Omodei} \cite{o:1}, Figure
5 (on page 87) with the corresponding $Re = 1$ result depicted in
Figure \ref{131}, for example. Good agreement is obtained between {\sc
Omodei's} Figure 7 (on page 88) result and the corresponding
$Re = 10$ result depicted in Figure \ref{132}, likewise between his
Figure 8 (on page 89) result and the corresponding $Re = 18$ result
depicted in Figure \ref{64}. These results are further substantiated by
those of {\sc Kruyt, Cuvelier, Segal} and {\sc Van Der Zanden}
\cite{k:1}, and those of {\sc Engelman} and {\sc Dupret} quoted in {\sc
Kruyt et al.}. \footnotetext[1]{Care needs to be taken in
matching inlet velocity profiles as well as scales when comparing
results.}

\subsection{Example 4: A Stokes Second Order Wave}

In this test the velocity profile and surface elevation predicted by
Stokes second order wave theory were used as boundary conditions for flow and free surface subproblems respectively. The
development of Stokes second order theory is the same as that of Airy
wave (Stokes first order) theory, the exception being that terms as far
as second order ($O \left[ \frac{H^2}{\lambda^2} \right]$) in the power
series expansion of the velocity potential are used. Wave troughs are
consequently flatter and the crests are sharper than those predicted
using the corresponding first order theory (see {\sc Koutitas}
\cite{koutitas:1} in this regard). 

\subsubsection*{Results}

\begin{figure}[H]
\begin{center} \leavevmode
\mbox{\epsfbox{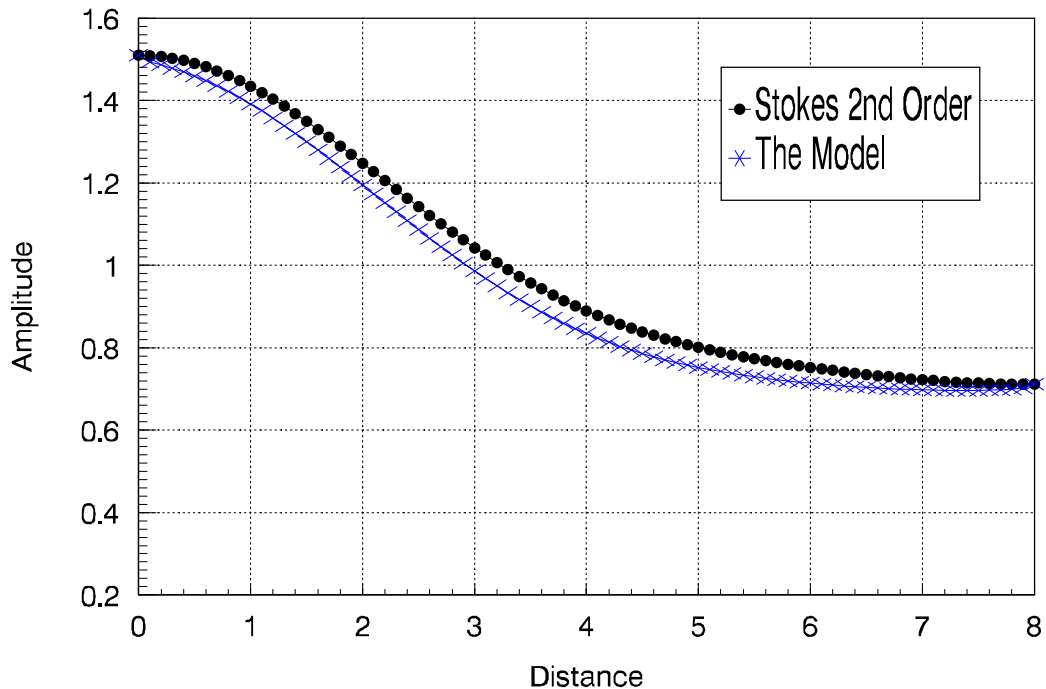}}
\end{center}
\caption{A Stokes Second Order Wave.}
\label{87}
\end{figure}

The method converged rapidly to the Stokes second order surface (Figure
\ref{87} on page \pageref{87}) in the initial stages, despite the use
of some fairly preposterous free surface starting guesses. This was
found to be so for a range of wave lengths, amplitudes and domains
experimented with. The problem was, however, not attempted with the
same seriousness as previous examples and the mesh was poor (there
being only three elements in the vertical extent of the mesh). Problems
with wave propagation were consequently experienced as time
progressed.

\section{Conclusions}

The completely general referential description proposed in {\sc Childs}
\cite{me:1} is varified by the numerical experiments of Section
\ref{4}, as was the predictor--corrector approach proposed in Section
\ref{5}. Although the resources at the authors disposal limited the
tests to two dimensional problems, this is not expected to be of any
consequence.

The linearised terms $( 2 {\bv} \mid_t - {\bv} \mid_{t-\Delta t} )
\cdot \nabla {\bv} \mid_{t+\Delta t} $ and $ {\bv} \mid_{t+\Delta t}
\cdot \nabla (2 {\bv} \mid_t - {\bv} \mid_{t-\Delta t})$ are second
order accurate approximations of the convective term. The formula given
in Theorem \ref{8} provides an exact quantitative measure of local
convergence for a given Reynolds number when applying a continuation
technique and, depending on the solver used, can improve efficiency.

The more efficient and more easily implemented penalty method was found
to produce comparable results to the iterative augmented Lagrangian
approach (through which incompressibility can be exactly enforced) in
the simple test examples performed. With regard to the L.B.B.
condition, it is important to note that a linear basis function mapped
from the master element using a $Q_2$ mapping will no longer be $P_1$
for non--rectangular elements (the $Q_2$--$P_1$ element pair was shown
to satisfy the L.B.B. condition in the context of rectangular
elements).

The holistic method to automatically generate meshes about rigid bodies
included within the fluid, using finite element mappings, was found to
be remarkably practical, simple and effective with maximum angles
within the mesh never exceeding $\frac{\pi}{4}$ radians (outlined in
Section \ref{9}).

\bibliography{cmame2}

\begin{thebibliography}{10}

\bibitem{b:4}
R.~B. Bird, R.~C. Armstrong, and O.~Hassager.
\newblock Dynamics of polymeric liquids.
\newblock {\em Fluid Mechanics, 2nd edn, Wiley, New York}, 1, 1987.

\bibitem{c:1}
G.~F. Carey and R.~Krishnan.
\newblock Penalty approximation of {S}tokes flow.
\newblock {\em Computer Methods in Applied Mechanics and Engineering},
  35:169--206, 1981.

\bibitem{me:1}
S.~J. Childs.
\newblock The energetic implications of using deforming reference descriptions
  to simulate the motion of viscous incompressible fluids.
\newblock {\em in review, International Journal of Numerical Methods in
  Fluids}, 1998.

\bibitem{me:3}
S.~J. Childs and B.~D. Reddy.
\newblock Finite element simulation of a rigid body in a fluid with free
  surface.
\newblock {\em Proceedings of the 1st South African Conference on Applied
  Mechanics '96}, pages 190--197, 1996.

\bibitem{c:3}
C.~Cuvelier, A.~Segal, and A.~A. van Steenhoven.
\newblock {\em Finite Element Methods and {N}avier--{S}tokes Equations}.
\newblock D. Reidel Publishing Company, Dordrecht, Holland, 1986.

\bibitem{h:3}
Thomas J.~R. Hughes.
\newblock Recent progress in the development and understanding of {SUPG}
  methods with special reference to the compressible {E}uler and
  {N}avier--{S}tokes equations.
\newblock {\em Computer Methods in Applied Mechanics and Engineering},
  7:1261--1275, 1987.

\bibitem{h:4}
Thomas J.~R. Hughes, Michel Mallet, and Akira Mizukami.
\newblock A new finite element formulation for computational fluid dynamics:
  {II}. {B}eyond {SUPG}.
\newblock {\em Computer Methods in Applied Mechanics and Engineering},
  54:341--355, 1985.

\bibitem{j:1}
Claes Johnson.
\newblock {\em Numerical Solution of Partial Differential Equations by the
  Finite Element Method}.
\newblock Cambridge University Press, 1987.

\bibitem{j:2}
D.~D. Joseph and Y.~Y. Renardy.
\newblock Fundamentals of two--fluid dynamics, part {II}: {L}ubricated
  transport, drops and miscible liquids.
\newblock {\em Springer, New York}, 1993.

\bibitem{koutitas:1}
Christopher~G. Koutitas.
\newblock {\em Mathematical Models in Coastal Engineering}.
\newblock Applied Mathematical Sciences. Pentech Press.

\bibitem{k:1}
N.~P. Kruyt, C.~Cuvelier, A.~Segal, and J.~Van~Der Zanden.
\newblock A total linearisation method for solving viscous free boundary flow
  problems by the finite element method.
\newblock {\em International Journal for Numerical Methods in Fluids},
  8:351--363, 1988.

\bibitem{l:1}
L.~D. Landau and E.~M. Lifshitz.
\newblock {\em Fluid Mechanics}.
\newblock Pergamon, Oxford, 1987.

\bibitem{oden:1}
J.~Tinsley Oden, Noboru Kikuchi, and Young~Joon Song.
\newblock Penalty--finite element methods for the analysis of {S}tokesian
  flows.
\newblock {\em Computer Methods in Applied Mechanics and Engineering},
  31:297--329, 1982.

\bibitem{o:1}
Bernard~J. Omodei.
\newblock Computer solutions of a plane {N}ewtonian jet with surface tension.
\newblock {\em Computers and Fluids}, 7:79--96, 1979.

\bibitem{s:1}
J.~C. Simo and F.~Armero.
\newblock Unconditional stability and long--term behaviour of transient
  algorithms for the incompressible {N}avier--{S}tokes and {E}uler equations.
\newblock {\em Computer Methods in Applied Mechanics and Engineering},
  111:111--154, 1993.

\bibitem{w:1}
N.~M.~J. Woodhouse.
\newblock {\em Introduction to Analytical Dynamics}.
\newblock Oxford Science Publications, 1987.

\end{thebibliography}

\end{document}